\documentclass[fleqn,usenatbib]{mnras}
\usepackage{newtxtext,newtxmath}
\usepackage[T1]{fontenc}
\usepackage{ae,aecompl}
\usepackage{times}
\usepackage{xcolor}

\usepackage{graphicx}	
\graphicspath{{/Users/chessipearce/Documents/PhD/paper/final_submission/}}
\usepackage{amsmath}	
\usepackage{amssymb}	
\usepackage{comment}
\usepackage{etoolbox}
\makeatletter
\patchcmd\@combinedblfloats{\box\@outputbox}{\unvbox\@outputbox}{}{%
	\errmessage{\noexpand\@combinedblfloats could not be patched}%
}%
\makeatother

\newcommand{\msol}{M$_{\odot}$}
\newcommand{\kb}{k_{\mathrm{B}}}
\newcommand{\bx}{b_{\mathrm{X}}}

\title[Massive cluster properties \& mass estimates]{Hydrostatic mass estimates of massive galaxy clusters: a study with varying hydrodynamics flavours and non-thermal pressure support}

\author[F. A. Pearce et al.]{Francesca A. Pearce $^{1}$,\thanks{E-mail: francesca.pearce@manchester.ac.uk}
Scott T. Kay $^{1}$, David J. Barnes $^{1,2}$, Richard G. Bower $^{3}$ and \newauthor Matthieu Schaller $^{4}$
\\
$^{1}$Jodrell Bank Centre for Astrophysics, School of Physics and Astronomy, The University of Manchester, Manchester M13 9PL, UK\\
$^{2}$Departmentof Physics, Kavli Institute for Astrophysics and Space Research, Massachusetts Institute of Technology, Cambridge, MA 02139, USA\\
$^{3}$Institute for Computational Cosmology, Department of Physics, University of Durham, South Road, Durham DH1 3LE, UK\\
$^{4}$Leiden Observatory, Leiden University, PO Box 9513, 2300 RA Leiden, the Netherlands
}

\date{Accepted XXX. Received YYY; in original form ZZZ}

\pubyear{2019}

\begin{document}
\label{firstpage}
\pagerange{\pageref{firstpage}--\pageref{lastpage}}
\maketitle

\begin{abstract}
	We use a set of 45 simulated clusters with a wide mass range ($8\times 10^{13} < M_{500}~[$M$_{\odot}]~< 2\times 10^{15}$) to investigate the effect of varying hydrodynamics flavours on cluster mass estimates. The cluster zooms were simulated using the same cosmological models as the BAHAMAS and C-EAGLE projects, leading to differences in both the hydrodynamic solvers and the subgrid physics but still producing clusters which broadly match observations. At the same mass resolution as BAHAMAS, for the most massive clusters ($M_{500} > 10^{15}$~M$_{\odot}$), we find changes in the SPH method produce the greatest differences in the final halo, while the subgrid models dominate at lower mass. By calculating the mass of all of the clusters using different permutations of the pressure, temperature and density profiles, created with either the true simulated data or mock spectroscopic data, we find that the spectroscopic temperature causes a bias in the hydrostatic mass estimates which increases with the mass of the cluster, regardless of the SPH flavour used. For the most massive clusters, the estimated mass of the cluster using spectroscopic density and temperature profiles is found to be as low as 50 per cent of the true mass compared to $\sim$ 90 per cent for low mass clusters. When including a correction for non-thermal pressure, the spectroscopic hydrostatic mass estimates are less biased on average and the mass dependence of the bias is reduced, although the scatter in the measurements does increase.
\end{abstract}




\section{Introduction}

Galaxy clusters are the rarest collapsed objects in the Universe, forming from the largest fluctuations in the primordial density field. A key result is that the abundance of clusters at any epoch depends sensitively on the cosmological initial conditions. Therefore, clusters are important cosmological probes, providing particularly stringent constraints on the matter density of the Universe, $\Omega_{\mathrm{m}}$, the amplitude of the matter power spectrum, $\sigma_{8}$, and the nature and evolution of dark energy (see \citealt{Voit2005a, Allen2011, Weinberg2013} for recent reviews).

Recently, groups working with data from the Atacama Cosmology Telescope (ACT, \citealt{Swetz2011, Hasselfield2013}), the South Pole Telescope (SPT, \citealt{Carlstrom2011, Benson2013, Reichardt2013}) and the \textit{Planck} satellite \citep{Tauber2010, PlanckI2011} have all used the Sunyaev-Zel'dovitch (SZ) effect (\citealt{Sunyaev1970}; \citeyear{Sunyaev1972}) to probe galaxy clusters. The cosmological parameters derived using SZ number counts from \textit{Planck} may be in tension with those derived using the cosmic microwave background (CMB), which, among other things, may be attributed to the uncertainty in measuring cluster masses \citep{PlanckXX2014, PlanckXXIV2016}. However, using SPT SZ cluster counts with updated weak-lensing calibration, the uncertainty of the observable-mass relation increases such that there is no tension with the \textit{Planck} CMB derived cosmological parameters \citep{deHaan2016}. In order to fully constrain cosmological parameters using galaxy clusters it is crucial to understand the cluster mass function and its evolution with redshift. Therefore, cluster masses need to be accurately calibrated against the observable signal and any measurement bias constrained.

Cluster masses can be estimated observationally using a variety of different techniques including X-ray observations, lensing and galaxy kinematics. In the first case, density and temperature profiles obtained from X-ray data are combined assuming hydrostatic equilibrium (e.g. \citealt{Vikhlinin2006}). However, many groups have shown that the assumption of galaxy clusters being in hydrostatic equilibrium leads to a bias in the estimated mass as a result of temperature inhomogenities, and turbulence and bulk motions as a result of mergers (e.g. \citealt{Nagai2007a, Rasia2012, Nelson2012, Khedekar2013, Martizzi2016, Shi2016}). Weak lensing mass estimates are obtained by fitting the shear profile of the cluster with a model profile from which the mass can be inferred. Studies of dark matter only simulations and those including baryons have both found that weak lensing mass estimates are biased low by $\sim~5-10$~per cent \citep{Becker2011, Bahe2012, Henson2017}. The SZ effect provides a way of obtaining the pressure profile of a cluster as the SZ signal, $y$, is proportional to the integrated pressure along the line-of-sight. This SZ pressure profile can then be combined with X-ray density to provide an alternative method of estimating the mass of a cluster. However, typically the SZ mass of a cluster is calculated through calibration of the $Y-M$ scaling relation (e.g. \citealt{deHaan2016}), where $Y$ is the integrated SZ signal and $M$ is the cluster mass estimated by other means.

From simulations, cluster masses found when assuming that clusters are in hydrostatic equilibrium are underestimated by around 20 per cent (e.g. \citealt{Nagai2007b, Kay2012, Rasia2012, LeBrun2014, Henson2017, Barnes2017a}). As clusters grow by hierarchical mergers, the intracluster medium (ICM) is heated through the release of gravitational energy. The kinetic energy of the infalling gas is expected to dissipate into heat so that the plasma is thermalised (e.g. \citealt{Kravtsov2012}), however the timescale over which kinetic energy is dissipated is currently unknown. The non-thermal processes which result from bulk motions and turbulence are predicted by simulations to contribute as much as 30 per cent to the overall pressure support in the outskirts of clusters today \citep{Nelson2014a}. In cluster cores, the total contribution from non-thermal pressure is thought to be much lower, at a level of only a few per cent, as was shown by the \textit{Hitomi} spacecraft observations of the core of the Perseus cluster \citep{Hitomi2016}. Magnetic fields and cosmic rays are also thought to contribute, but only on the level of a few per cent (e.g. \citealt{Ackermann2014, Brunetti2014}), and are not currently included in our cosmological simulations. Another physical process missing from most simulations is thermal conduction which impacts the fraction of cool-core clusters and increases the efficiency of feedback from black holes \citep{Yang2016, Barnes2018b}.

For simulations of the most massive clusters, $M~>~10^{15}$~\msol, \cite{Henson2017} found that the bias between the estimated and true mass of a simulated increased to around 40 per cent from 20 per cent for clusters with $M \approxeq 10^{14}$~\msol~(see also \citealt{Barnes2017b}). This bias was found to be due to the presence of cooler gas permeating through the clusters which resulted in a biased temperature profile constructed from mock spectroscopic data (using the method of \citealt{LeBrun2014, Barnes2017a, Barnes2018c}). Similarly, \cite{Rasia2012} found that the average X-ray mass bias in a sample of simulated clusters was 25-35 per cent due to temperature inhomogeneties, but no comment about a potential mass dependence of this bias was made.

It is known that the solutions to the smoothed particle hydrodynamics (SPH) equations which underpin the cosmological simulations used by \cite{Henson2017} (henceforth \textsc{gadget}-SPH) lead to a lack of mixing \citep{Mitchell2009,Read2010,Sembolini2016a} which could cause this bias. Newer SPH flavours aim to increase the mixing of gas by altering the SPH equations (e.g. \citealt{Ritchie2001, Read2012,Hopkins2013,Price2017}), and/or by adding an additional term which diffuses the entropy of particles across discontinuities \citep{Price2008}.

In this paper, we study the hydrostatic mass bias using a sample of 45 clusters that were simulated using both \textsc{gadget}-SPH and a newer flavour that includes a switch for artificial conduction to improve the mixing between gas phases \citep{Price2008}. Specifically, we aim to test whether the large bias found by \cite{Henson2017} is reduced when the newer scheme is adopted. A secondary aim is to investigate how cluster mass estimates are affected when applying a non-thermal pressure correction. The rest of the paper is organised as follows. In Section~\ref{Sec:Simulations}, the cluster sample is introduced and the difference between the models used to simulate the clusters is explained. In Section~\ref{Sec:GasProperties}, we compare the hot gas properties of the clusters, and in Section~\ref{Sec:Mass} we show results for the hydrostatic masses of the clusters and determine whether a correction for non-thermal pressure reduces the mass bias. Our conclusions are then given in Section~\ref{Sec:Conclusions}.

\section{Simulations \& analysis pipeline}\label{Sec:Simulations}

\subsection{Cluster sample}
In this paper we use a sample of 45 objects which includes all 30 C-EAGLE clusters \citep{Barnes2017b} and an additional 15 higher-mass clusters chosen from the same parent simulation (described in \citealt{Barnes2017a}). These extra clusters were included to increase the statistics at high mass to investigate any potential mass dependence of the hydrostatic mass bias. The overall mass range was $8\times 10^{13} < M_{500}$\footnote{We define $M_{500}$ as the mass of a cluster within a sphere of radius $r_{500}$ that encloses a density equal to 500 times the critical density of the Universe, $\rho_{\mathrm{crit}}$.}$~[$M$_{\odot}] < 2\times 10^{15}$, similar to that of \cite{Henson2017}. All simulations were carried out at the same mass resolution as the objects in the BAHAMAS project \citep{McCarthy2017}. This resolution is considerably lower than the C-EAGLE runs ($m_{\mathrm{EAGLE}}/m_{\mathrm{BAHAMAS}} \simeq 1.5\times 10^{-3}$), hence we refer to the set of 45 clusters as the `C-EAGLE Low Resolution' (CELR) sample. Where a specific cluster is discussed, it is referred to as CE-X where X is the same halo number as in the C-EAGLE analysis\footnote{With the exception of halo 40 in this analysis which is halo 29 in the C-EAGLE sample.}  \citep{Barnes2017b}.

The 45 clusters were simulated three times, each with varying SPH flavour and/or subgrid model, as detailed below. For each cluster the same initial conditions were used regardless of the simulaton model. We acknowledge that this is not a complete, robust study into the effect of SPH flavour on the hydrostatic mass bias due to not systematically changing the SPH flavour without editing the subgrid models.

All of the simulations utilised the zoom technique of \cite{Katz1993}, where the resolution of the area of interest was increased relative to the surroundings whilst using lower resolution particles to maintain the correct tidal forces around the cluster (see also \citealt{Tormen1997}). The dark matter particles had a mass of $m_{\mathrm{DM}} = 6.5\times 10^{9}$~\msol and the gas particles had an initial mass of $m_{\mathrm{gas}} = 1.2\times 10^{9}$~\msol. The gravitational softening was set to 4 kpc/$h$ in physical coordinates below $z=3$ and fixed to the same value in co-moving coordinates at higher redshift. The minimum SPH smoothing length was fixed at 0.1 times the gravitational softening length at all times. The cosmological parameters used in the simulations were set to those used in EAGLE \citep{Schaye2015, Crain2015}, based on a flat $\Lambda$CDM universe combining the \textit{Planck} 2013 results with baryonic acoustic oscillations, \textit{WMAP} polarization and high multipole experiments \citep{PlanckXVI2014}; $\Omega_{\mathrm{b}} = 0.04825$, $\Omega_{\mathrm{m}} = 0.307$, $\Omega_{\Lambda} = 0.693$, $h \equiv H_{0}/(100~\mathrm{kms}^{-1}~\mathrm{Mpc}^{-1}) = 0.6777$, $\sigma_{8} = 0.8288$, $n_{\mathrm{s}} = 0.9611$ and $Y = 0.248$. These parameters vary slightly from those used in the BAHAMAS simulations which were based only on \textit{Planck} 2013 results \citep{PlanckXVI2014}. Both the paramater sets we could have chosen are consistent with the \textit{Planck} cosmology so the results presented here would not have been affected by changing to the (slightly different) BAHAMAS cosmological parameters (e.g. \citealt{Henson2017,McCarthy2017}).

Once the individual clusters were simulated, \textsc{subfind} \citep{Springel2001, Dolag2009} was run to identify the main halo and any sub-haloes. During the zoom simulation, an on-the-fly friends-of-friends (FOF) algorithm was run that linked all particles within a dimensionless linking length, $b=0.2$, which can lead to the inclusion of non-gravitationally bound particles in the haloes \citep{Davis1985}. \textsc{subfind} refines the FOF groups by identifying the most bound particle (which defines the centre of the group) and removing any particles that are not gravitationally bound to distinct sub-haloes.

\subsubsection{BAHAMAS runs}
We first simulated the 45 clusters using the same cosmological model as used for the BAHAMAS project \citep{McCarthy2017} and the MACSIS higher-mass extension \citep{Barnes2017a}. These runs, labelled CELR-B, utilise a heavily modified version of \textsc{gagdet-3} (last described as \textsc{gadget-2} by \citealt{Springel2005b}) that includes the subgrid physics developed as part of the OWLS project \citep{Schaye2010}. The underlying SPH flavour of the BAHAMAS model, \textsc{gadget}-SPH, is described in detail in \cite{Springel2002}. The main features are the smoothing function for gas particles which uses the $M_4$ cubic B-spline function from which the smoothing length is calculated by interpolating the 48 nearest neighbour particles. Particle velocities and positions are integrated in time along with the entropic function, which relates a particle's pressure and density. In order to increase the entropic function in the presence of shocks, artificial viscosity is implemented in the simulation by adding a term to the equations of motion. A viscous tensor is used to represent an additional pressure with viscosity parameter $\alpha$, while a shear flow switch $f_i$ prevents the application of viscosity where there are pure shear flows \citep{Balsara1995}.

A full description of the subgrid physics can be found in e.g. \cite{Schaye2010}, \cite{LeBrun2014}, \cite{McCarthy2017}. Briefly, the main features are radiative cooling which is calculated on an element-by-element basis following the model of \cite{Wiersma2009a};  the formation of stars at a pressure-dependent rate by gas particles with density $n_{\text{H}} > 0.1\text{ cm}^{-3}$ which obey an equation of state $P \propto \rho^{4/3}$, where $P$ is the gas pressure and $\rho$ is the gas density \citep{Schaye2008}; stellar evolution and enrichment using the model of \cite{Wiersma2009b} which follows 11 chemical elements, and assumes that winds, Type Ia and Type II supernovae drive mass loss; a stellar feedback model \citep{Dalla2008}; and supermassive black hole seeding, growth and feedback models \citep{Booth2009}.

The kinetic stellar feedback model of \cite{Dalla2008} was used due to the difficulty in simulating thermal feedback at this resolution as typically the thermal energy is radiated away too quickly leading to overcooling (e.g. \citealt{Katz1996, Balogh2001}). For each star particle $j$, the probability of its SPH neighbour $i$ receiving a kick of velocity $v_{\mathrm{w}}$ is given by $\eta m_{j}/\Sigma_{i=1}^{N}m_{i}$. If all baryon particles had equal mass then each star particle would kick on average $\eta$ particles. The values used of $\eta = 2$ and $v_{\mathrm{w}} = 600$ km s$^{-1}$ correspond to winds carrying approximately 40 per cent of the energy available from core collapse supernovae assuming a \cite{Chabrier2003} initial mass function. Contrary to other kinetic feedback models, the winds (kicked particles) are not hydrodynamically decoupled from the surrounding gas. For more information on calibration and comparisons to observations, see \cite{Schaye2010} and \cite{McCarthy2017}.

Feedback from active galactic nuclei (AGN) has been shown to be crucial in producing realistic clusters (e.g. \citealt{Borgani2011, Planelles2013, Planelles2014, LeBrun2014, Pike2014, Sembolini2016b, Barnes2017a, McCarthy2017, Planelles2017}). However, the physics of the formation of black holes (BHs) is currently not well understood. Instead, BH seed particles are placed in the centre of overdense regions in the simulations. The model of \cite{Booth2009}, a modification of that in \cite{Springel2005a}, includes the use of an on-the-fly FoF algorithm which enables BH seed particles to be placed in haloes with at least 100 DM particles (corresponding to $\sim 6.5\times 10^{11}\text{ M}_{\odot}$). The black holes can then grow via mergers or the accretion of surrounding gas, where the accretion rate is equal to a scaled up Bondi-Hoyle-Lyttleton rate. The scale factor, $\alpha$, is a power-law function of the local density such that
\begin{equation}\label{eq:alpha}
\alpha = 
\begin{cases}
1 & \mathrm{if}~n_{\mathrm{H}} < n^{*}_{\mathrm{H}},\\
\left(n_{\mathrm{H}}/n^{*}_{\mathrm{H}}\right)^{\beta} & \mathrm{otherwise},
\end{cases}
\end{equation}
where $n^{*}_{\mathrm{H}} = 0.1$~cm$^{-3}$ is the critical density required for the formation of a cold interstellar gas phase. Here, $\alpha = 1$ for densities equal to the star formation threshold (or when $\beta=0$) so the accretion rate is equal to the Bondi-Hoyle-Lyttleton rate in this regime. A self-regulating feedback mechanism is established as a fraction $\epsilon=0.015$ of the rest mass energy of the accreted gas which is used to heat $n_{\text{heat}} = 20$ nearest neighbour particles by increasing their temperature by $\Delta T_{\text{heat}} = 10^{7.8}\text{ K}$.

\subsubsection{EAGLE runs}
A second set of runs, labelled CELR-E, use the same cosmological model as for the EAGLE project. It is again a highly modified version of \textsc{gadget-3} whose subgrid model is based on OWLS. Of central importance to this study is that the EAGLE model has improved hydrodynamic algorithms known as \textsc{anarchy}, described in \citeauthor{Schaye2015} (\citeyear{Schaye2015}, Appendix A) and \cite{Schaller2015}. The smoothing lengths are calculated using the $C_2$ kernel \citep{Wendland1995} and 58 nearest neighbour particles, which yields the same full width at half-maximum as \textsc{gadget} \citep{Dehnen2012}. The equations of motion are based on the pressure-entropy formulation of \cite{Hopkins2013}, in which the density, velocity and entropic function of particles are integrated in time along with two new quantities, the weighted density and associated weighted pressure. These new quantities act to smooth the pressure at contact discontinuities. To reduce the viscosity away from shocks, the viscosity parameter $\alpha$ is modelled as a differential equation for each particle instead of a constant (e.g. \citealt{Morris1997,Cullen2010}). The final difference in the SPH formulations is the addition of entropy diffusion between particles, following \cite{Price2008}. This leads to better mixing of entropies and the creation of a single-phase gas at discontinuities (e.g. \citealt{Read2012,Hopkins2013}).

The subgrid model for EAGLE is described in \cite{Schaye2015} (see also \citealt{Crain2015} for details of the calibration and impact of changing the model). We did not recalibrate the model for the mass resolution used in this analysis as our main goal was to study the effects of varying the SPH flavour. The main difference between the models used in EAGLE and BAHAMAS is related to the AGN and stellar feedback prescriptions.

In EAGLE, black holes are seeded at the centres of FoF haloes (defined on-the-fly) with a minimum of 32 DM particles and total mass greater than $10^{10}\text{ M}_{\odot}/h$, by converting the most dense gas particle to a BH seed particle with $m_{\text{BH}} = 10^5\text{ M}_{\odot}/h$. Due to the decrease in mass resolution in our simulations compared to the EAGLE runs, the minimum halo mass required for a BH seed particle is increased to $\sim~2\times 10^{11}$~\msol, equivalent to 32 DM particles. The BH can then grow via accretion or mergers as in the BAHAMAS model, but the gas accretion rate depends on the effective viscosity of the accretion disc and does not use a density dependent boost factor due to the assumed resolution of the simulation \citep{Rosas2015}. The AGNdT9 feedback model was used in the C-EAGLE project as it has been shown to produce a good match to the observed gas mass fractions of low mass groups \citep{Schaye2015}. Although the overall energy released per feedback event is roughly the same between EAGLE and BAHAMAS (within $\sim~25$~per cent) for equal mass particles, the injection method between the two models is different. In EAGLE, instead of heating 20 particles by $\Delta T = 10^{7.8}\text{ K}$ when the BH energy is above the corresponding threshold, thermal energy is instead injected stochastically. Above a critical energy the BH has a probability $P$ of heating a single particle by an amount $\Delta T = 10^9\text{ K}$. If there is still energy left above the threshold after a heating event, the timestep of the BH is shortened so that more energy can be released.

Instead of implementing stellar feedback using a kinetic model, EAGLE uses the stochastic thermal feedback model of \cite{Dalla2012}. The temperature jump of a heated particle is defined as $\Delta T = 10^{7.5}$ K, and the probability that an SPH neighbour of the stellar particle will be heated is related to the fraction of the total energy per core collapse supernova which is injected per unit stellar mass on average, $f_{\mathrm{th}}$. These heating events happen once per stellar particle when it has reached $3\times 10^{7}$ yr (the maximum lifetime of a star which explodes as a core collapse supernova). For more detailed information see \cite{Schaye2015}. Contrary to other stellar feedback models, the EAGLE model does not assume any wind velocity or mass loading properties. As this model was chosen assuming higher mass resolution than the simulations carried out in this analysis, it is not expected that the stellar feedback will be optimal in the new runs.

Changes were also made to the reionization and star formation models. Instead of being switched on at $z=9$, reionization occurs at $z=11.5$ for hydrogen with an injection of 2 eV per proton mass and is Gaussian distributed for helium with a width of $\sigma(z)=0.5$ centred on $z=3.5$. For star formation, the density threshold for particles which can form stars is dependent on metallicity $Z$ in the EAGLE model; $n_{\text{H}} > 0.1\text{ cm}^{-3}(Z/0.002)^{-0.64}$ \citep{Schaye2004}.

For the final set of runs, labelled CELR-NC, we use the same model as for CELR-E but with artificial conduction turned off. This was done because tests with non-radiative hydrodynamical simulations have shown that artificial conduction, a new feature in the \textsc{anarchy} flavour of SPH, has the largest effect on the entropy in cluster cores due to mixing. It is again worth stressing that despite changing the mass resolution we did not recalibrate the EAGLE model for CELR-E or CELR-NC. Section~\ref{Sec:GasProperties} gives a detailed comparison of the final clusters for all three sets of runs and shows that the properties of the clusters are reasonable despite the lack of recalibration.

Due to changing the mass resolution of the simulations without then recalibrating the EAGLE models, there are a number of clusters within the CELR-E/NC samples with central BHs that are less massive than their CELR-B counterpart. When comparing the black hole masses to CELR-B there are two distinct groups of BHs, those with a median mass of nearly 20 per cent of the CELR-B sample, and those which are less. The BAHAMAS model includes a density dependent scale factor, $\alpha$ (Eq.~\ref{eq:alpha}), which is used to boost the accretion rate of BHs in low gas density regions such that the BH growth can become self-regulated \citep{Springel2005a}. Due to the assumed mass resolution of the simulations, this $\alpha$ factor is not included in the EAGLE model, leading to this dichotomy in the mass of CELR-E/NC BHs. Some of the clusters have central BHs which are unable to effectively accrete gas so do not reach the Edington regime and essentially do not grow above the seed mass. We repeated all the analysis presented in the paper with these two groups and found that none of the results changed significantly due to the size of the central BHs. However, we note that on average the CELR-E/NC clusters have central BHs that are at least a factor of 4 smaller than the comparable CELR-B cluster, so differences between the BAHAMAS and EAGLE based samples are still possible, especially as there are other changes to the subgrid models and different SPH flavours are used.

\subsection{Analysis pipeline}\label{Sec:Analysis}

In general, the clusters were analyzed in three different mass bins defined at $z=0$ to be:
\begin{itemize}
	\item $8.1\times 10^{13} < M_{500,\mathrm{true}}~[\mathrm{M}_{\odot}] < 3.0\times 10^{14}$,
	\item $3.0\times 10^{14} < M_{500,\mathrm{true}}~[\mathrm{M}_{\odot}] < 1.23\times 10^{15}$,
	\item $M_{500,\mathrm{true}} > 1.23\times 10^{15}~\mathrm{M}_{\odot}$,
\end{itemize}
where $M_{500,\mathrm{true}}$ is the true value of $M_{500}$, explained in more detail below. Throughout this paper, the mass bins will be referred to as the low, middle and high mass bins respectively. This definition meant that there were initially 15 clusters in each mass bin for the CELR-E run which resulted in a fluctuation of up to one cluster per mass bin for the other two simulation samples. The only exception to this definition is in Section~\ref{Sec:MassFractions} where we define mass bins with respect to the estimated spectroscopic masses (defined in Section~\ref{Sec:Spec}) to compare gas mass fractions to observed trends.

Clusters were defined as dynamically relaxed using the same criterion as \cite{Barnes2017b}, namely if
\begin{equation}\label{eq:relax}
E_{\mathrm{kin},500} < 0.1~E_{\mathrm{therm},500},
\end{equation}
where $E_{\mathrm{kin},500}$ is the sum of the kinetic energy of the gas particles within $r_{500}$, with the bulk motion of the cluster removed, and $E_{\mathrm{therm},500}$ is the sum of the thermal energy within the same radius. At $z=0$, 15 out of 45 clusters are classed as relaxed in the CELR-B sample, 14 for CELR-E, and 13 for CELR-NC. In all relevant figures shown below, relaxed (unrelaxed) clusters are represented by filled (open) points. 

\subsubsection{`True' vs. `spec' data}\label{Sec:Spec}
Throughout this analysis three-dimensional profiles are presented for both `true' and `spec' data. The former corresponds to using the particle data produced directly by simulations, whereas `spec' corresponds to mock data which is produced to more closely resemble X-ray observations. Inhomogeneities in the hot gas of observed clusters can cause biases in the ICM properties inferred from X-ray observations \citep{Nagai2007a, Khedekar2013}, so a comparison of the true and spec profiles show the impact of substructures and multi-temperature gas.

To create the true profiles, particles within each cluster were split into 50 radial bins, equally spaced between $(0.05-2)~r_{500,\mathrm{true}}$. The density within each radial bin was calculated as $\rho = \Sigma_{i}m_{i}/V_{\mathrm{shell}}$, where $m_{i}$ is the mass of the $i$th particle and $V_{\mathrm{shell}}$ is the volume of the shell. Similarly, the mass-weighted temperature $T_{\mathrm{mw}} = (\Sigma_{i}m_{i}T_{i})/\Sigma_{i}m_{i}$, where $T_{i}$ is the temperature of the $i$th particle. Particles with temperature $k_{\mathrm{B}}T < 10^{5}$~K or density $n \geq 0.1$~cm$^{-3}$ were removed from the profiles. 

Mock spectroscopic data was produced using the method of \citeauthor{LeBrun2014} (\citeyear{LeBrun2014}, see also \citealt{McCarthy2017, Barnes2017a, Barnes2017b, Barnes2018c}). For every gas particle with $T > 10^{5}$~K, a rest frame X-ray spectrum in the 0.5-10.0~keV band was produced using the Astrophysical Plasma Emission Code (\textsc{APEC}, \citeauthor{Smith2001} \citeyear{Smith2001}) via the \textsc{PyAtomDB} module with atomic data from AtomDB 3.0.8 \citep{Foster2012} which includes emission lines. Individual spectra were created for all of the 11 elements tracked by the simulations: H, He, C, N, O, Ne, Mg, Si, S, Ca and Fe. Particles were binned into 50 linearly-spaced radial bins and the summed spectra within each bin was scaled by the relative abundance of heavy elements as the fiducial spectra assume solar abundance \citep{Anders1989}. A single temperature APEC model was fitted to the spectrum in the range 0.5-10.0~keV to give an estimate of the temperature, density and metallicity in each radial bin. During fitting, the spectrum in each energy bin is multiplied by the effective area of \textit{Chandra} to provide a better match to X-ray observations. Our method does not fully reproduce X-ray observational analyses as we do not consider the effects of projecting the data; we leave such a study to future work.

For the hot gas density, $\rho$, temperature, $T$, and pressure, $P$, profiles presented in Section~\ref{Sec:GasProperties} (both true and spec), the expected self-similar mass and redsfhift dependence from gravitational heating can be removed by normalising each quantity respectively by
\begin{equation}\label{eq:rho}
\rho_{\mathrm{crit}}(z) \equiv E^2(z) \frac{3H_{0}^2}{8\pi G},
\end{equation}
\begin{equation}
T_{500} = \frac{GM_{500}\mu m_{\mathrm{p}}}{2k_{\mathrm{B}}r_{500}},
\end{equation}
\begin{equation}\label{eq:pressure}
P_{500} = 500f_{\mathrm{b}}k_{\mathrm{b}}T_{500} \frac{\rho_{\mathrm{crit}}}{\mu m_{\mathrm{p}}},
\end{equation}
where $E(z) \equiv H(z)/H_{0} = \sqrt{\Omega_{\mathrm{m}}(1+z)^{3}+\Omega_{\Lambda}}$, $H_{0}$ is the Hubble constant at $z=0$, $G$ is the gravitational constant, $k_{\mathrm{B}}$ is the Boltzmann constant, $\mu = 0.59$ is the mean molecular weight (assuming a primodial, fully ionzed plasma), $m_{\mathrm{p}}$ is the mass of a proton and $f_{\mathrm{b}} \equiv \Omega_{\mathrm{b}}/\Omega_{\mathrm{M}} = 0.157$ is the universal baryon fraction. The average of each normalization quantity was taken in each mass bin so that all the profiles are normalized by the same value. The profiles themselves are medians of the individual cluster profiles in each mass bin so that extreme clusters which are not representative of the overall sample were removed.

\begin{figure*}
	\includegraphics[width=\textwidth,trim=6cm 0.5cm 6cm 0.5cm, clip, keepaspectratio=True]{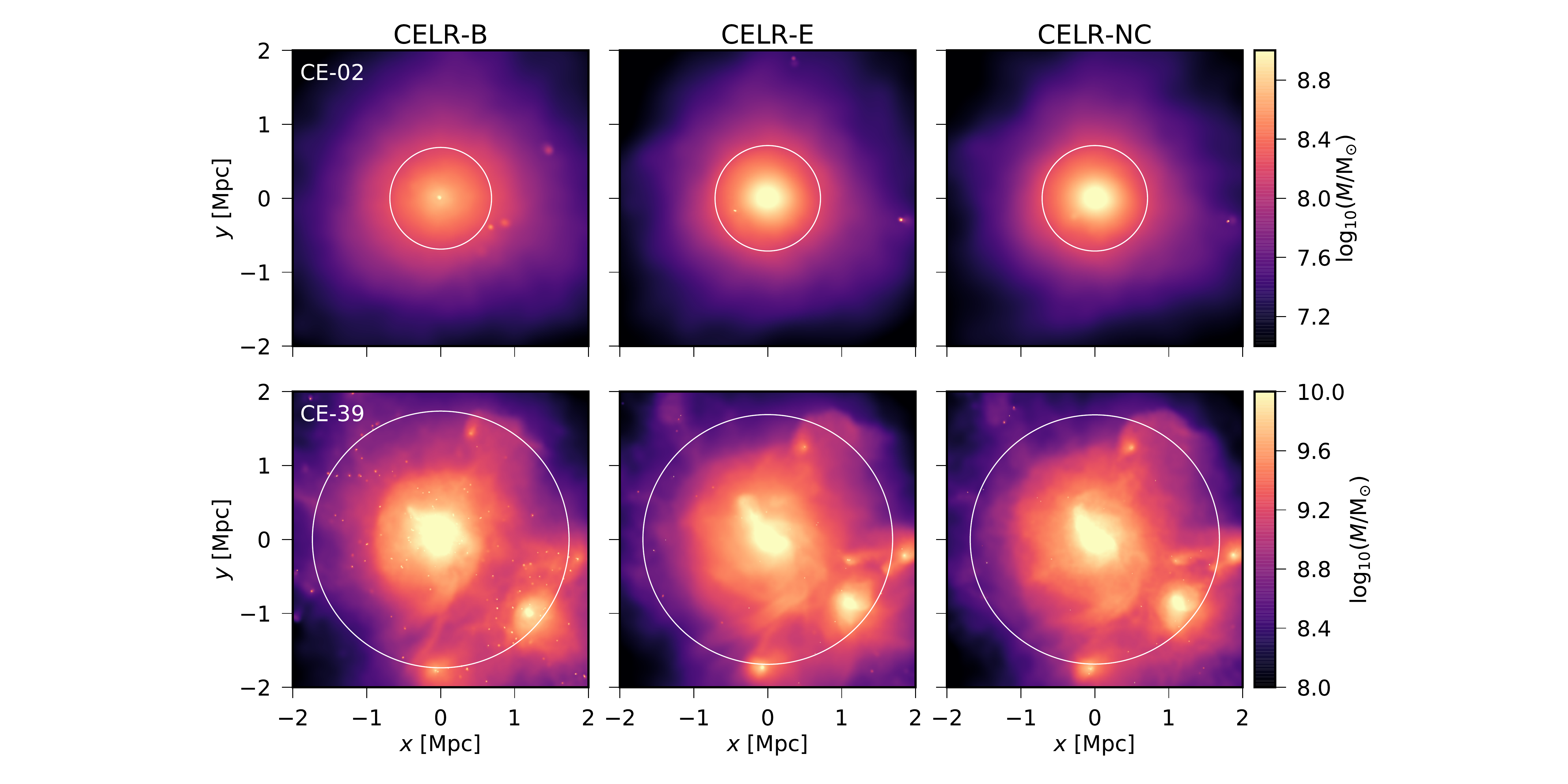}
	\caption{Projected gas mass maps of a relaxed cluster (CE-02, top row) and a dynamically active cluster (CE-39, bottom row). From left to right the maps show the resulting CELR-B, CELR-E and CELR-NC cluster. The white circle shows the position of $r_{500,\mathrm{true}}$. The biggest difference can be seen when comparing the CELR-B clusters to either the CELR-E or CELR-NC clusters as there are more differences in the underlying models than between CELR-E and CELR-NC. Improved mixing leads to a smoother mass distribution in the centre of CELR-E CE-39, while changes in the AGN feedback models cause the core of CELR-B CE-02 to be less massive compared to the other runs.}
	\label{fig:MassMap}
\end{figure*}

\begin{figure*}
	\includegraphics[width=\textwidth,trim=6cm 0.5cm 6cm 0.5cm, clip, keepaspectratio=True]{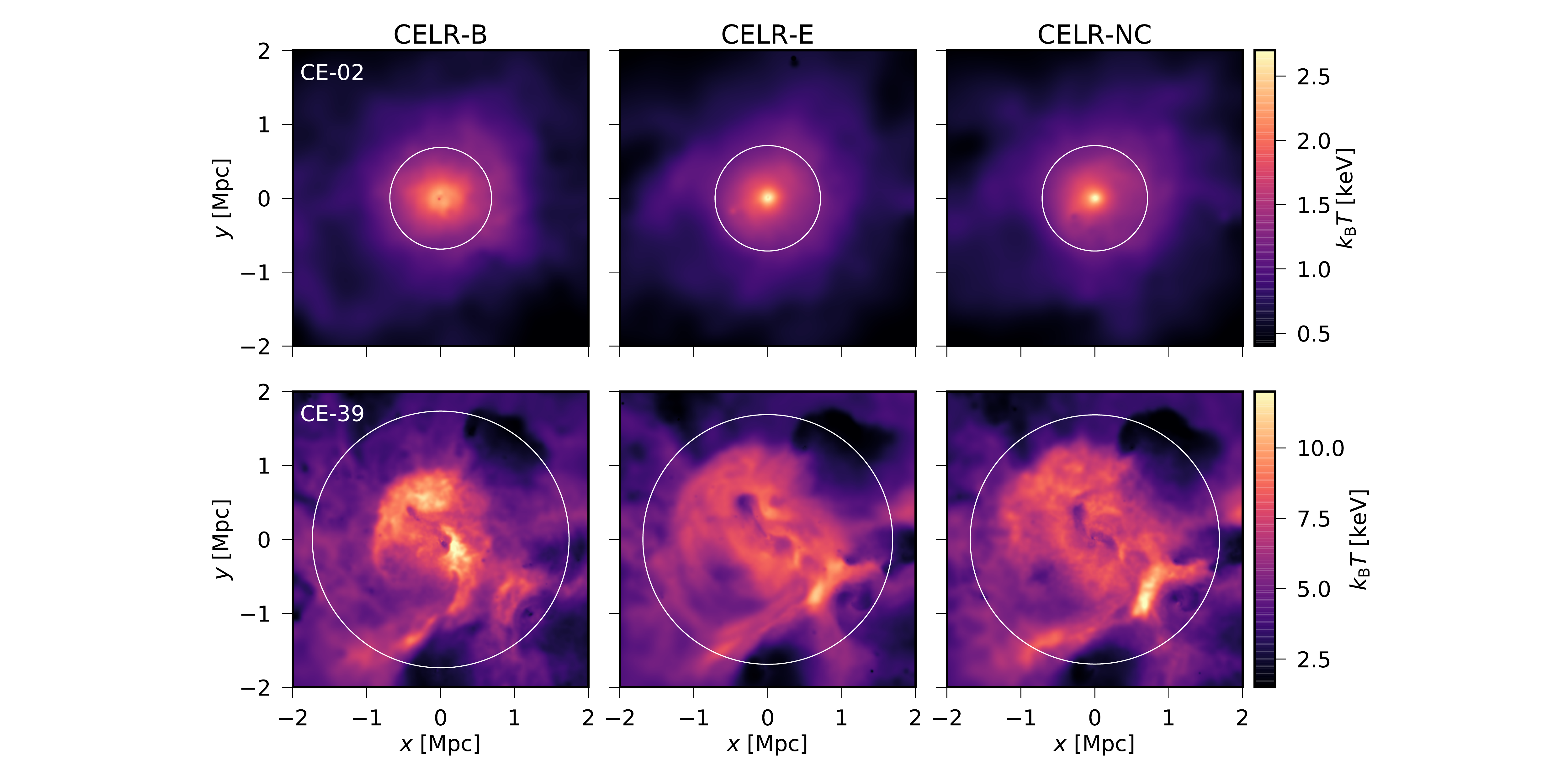}
	\caption{Projected mass-weighted temperature maps with the same layout as in Fig.~\ref{fig:MassMap}. Again the biggest differences between the final clusters can be seen when comparing CELR-B to CELR-E/NC, although the effect of artificial conduction can be seen in CELR-E CE-39 which shows a smoother temperature distribution compared to CELR-NC.}
	\label{fig:TempMap}
\end{figure*}

In general, profile data is presented between $(0.05-2.0)~r_{500}$ to show the cluster cores and the outskirts probed by the SZ effect. Where models have been fit to the data the fitting range adopted is (0.15-1.5)~$r_{500}$ which allows for the estimation of $r_{500}$ and $M_{500}$ from the spec profiles. The radius of convergence \citep{Power2003} of the smallest cluster was $0.004~r_{500,\mathrm{true}}$ for all three runs.

\section{Gas properties}\label{Sec:GasProperties}

In this section we discuss the hot gas properties of the clusters to try to understand the relative effects of changing the subgrid models and SPH flavour. First, we present the projected gas mass and mass-weighted temperature maps to qualitatively show the difference between clusters for different runs, and the baryonic content for the entire sample relative to observational trends. Then we examine the different profiles (hot gas density, temperature and pressure) which will be used to calculate a cluster's hydrostatic mass (Section~\ref{Sec:Mass}). In general, CELR-B results will be presented in red, CELR-E in blue and CELR-NC in green.

\subsection{Projected gas maps}\label{Sec:GasMaps}

Figs.~\ref{fig:MassMap} and \ref{fig:TempMap} show the projected gas mass and mass-weighted temperature maps respetively, for two different clusters from the CELR-B, CELR-E and CELR-NC runs (left to right panels respectively). All particles with $T > 10^{5}$~K and $n < 0.1$~cm$^{-3}$ within a cube of 4~Mpc are included in the map, centred on the most bound particle in the cluster. The top panels show CE-02 which is a relatively relaxed cluster $(E_{\mathrm{kin},500}/E_{\mathrm{therm},500} = 0.033\pm 0.002)$ and has $M_{500,\mathrm{true}} = (9.4^{+0.3}_{-0.7})\times 10^{13}$~\msol. The bottom panels are for CE-39 which is defined as unrelaxed due to the merging substructures present in the outskirts $(E_{\mathrm{kin},500}/E_{\mathrm{therm},500} = 0.18^{+0.01}_{-0.02})$, and has a higher mass, $M_{500,\mathrm{true}} = (1.3\pm 0.1)\times 10^{15}$~\msol. For each halo, the cluster properties, $E_{\mathrm{kin},500}/E_{\mathrm{therm},500}$ and $M_{500,\mathrm{true}}$, have been averaged over the three runs with the error giving the scatter in the values. The white circle shows the position of $r_{500,\mathrm{true}}$ for all the clusters.

Focussing on the smaller cluster (top row), the main differencs can be seen in the cluster centres. When looking at the mass maps (Fig.~\ref{fig:MassMap}), CELR-B CE-02 has a physically smaller core and less hot gas overall compared to either CELR-E or CELR-NC. This is likely attributed to the different feedback prescriptions. Previous work has shown that the majority of gas is released in the high redshift progenitors of massive clusters \citep{McCarthy2011}, so the model used in the BAHAMAS project is more efficient than that used in C-EAGLE at removing gas from the potential wells of the cluster progenitors at this resolution. CE-02 is defined as having a small central BH compared to CELR-B for both the CELR-E and CELR-NC samples due to the EAGLE model not boosting the accretion rate of BHs in low density regions. As feedback events only happen when the BH has reached a certain threshold to ensure there is enough energy available for heating, the CELR-E/NC clusters have had fewer feedback events as a result of the BHs growing more slowly.

Conversely, Fig.~\ref{fig:TempMap} shows that the very centre of CELR-B CE-02 has a lower temperature than either CELR-E or CELR-NC, so it is populated by colder, more dense gas. If this was caused by the lack of entropy diffusion in the BAHAMAS model, then it would follow that CELR-NC CE-02 would also have a colder core which is not seen. It is unclear whether the difference seen is due to different subgrid physics between the BAHAMAS and EAGLE models, other differences in the SPH (e.g. using pressure-entropy) or both. The main effects of improved mixing can, however, be seen in the more massive cluster, CE-39. Looking at the projected gas mass maps, Fig.~\ref{fig:MassMap}, the physical size of the core is comparable between the runs but there is a noticeable difference in the distribution of the central gas. At this mass AGN feedback is expected to be less important as the progenitors of more massive clusters, on average, collapse earlier when the Universe is more dense. As such, the progenitors have deeper potential wells from which the AGN can expel less gas. The artificial conduction included in ANARCHY allows for improved mixing of gas which results in fewer low entropy gas clumps falling into the centre of the final CELR-E cluster compared to CELR-B/NC. This can be seen for CELR-B CE-39, in which there are a number of small substructures throughout the cluster, especially towards the outskirts, that are the result of sinking low entropy gas that does not mix. This is similar to the results of \cite{Schaller2015} who compared \textsc{gadget}-SPH and ANARCHY using a cosmological box of side length 50~Mpc, run at EAGLE resolution. Looking at the projected mass-weighted temperature maps, Fig.~\ref{fig:TempMap}, the distribution of the gas is more smooth for CELR-E than in CELR-B or CELR-NC which do not include mixing. In general, we find subgrid prescriptions have less impact on the final cluster than changes to the SPH flavour for massive clusters.

\subsection{Gas \& stellar mass fractions}\label{Sec:MassFractions}

\begin{figure*}
	\includegraphics[scale=0.65,keepaspectratio=True]{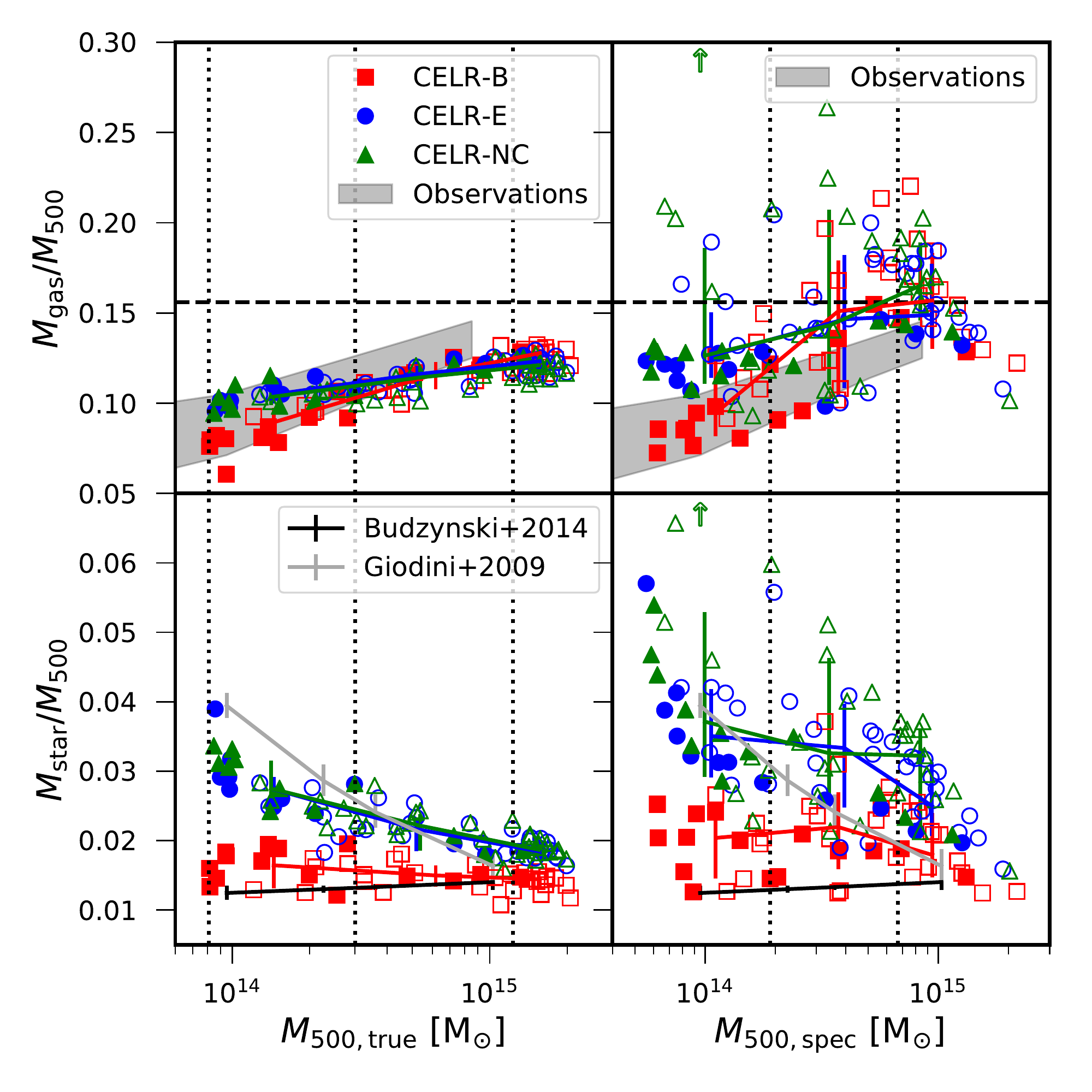}
	\caption{Gas (top row) and stellar (bottom row) mass fractions for individual clusters within $r_{500,\mathrm{spec}}$ versus true mass (left) and estimated spec mass (right). Red squares correspond to CELR-B, blue circles to CELR-E and green triangles to CELR-NC. Relaxed clusters are shown as filled points while the vertical dotted lines show the mass bins used to obtain the median coloured lines for each sample. If a cluster has a mass fraction which is outside of the plotted region it is instead displayed as an arrorw but with the correct colour and fillstyle of it's sample and relaxation state. For the gas mass fractions the grey band gives the 1~$\sigma$ scatter of combined observational samples, while the horizontal dashed black line represents the universal baryon fraction, $\Omega_{\mathrm{b}}/\Omega_{\mathrm{M}} = 0.157$. In the bottom panels the black line gives the relationship between stellar mass fraction and mass found by \citet{Budzynski2014}, while the grey line is the best fit found by \citet{Giodini2009}.}
	\label{fig:GasFractions}
\end{figure*}

The gas and stellar mass fractions of each cluster are presented in Fig.~\ref{fig:GasFractions} as the top and bottom panels respectively. The left column gives the true mass fractions found by summing the mass of all relevant particles within $r_{500,\mathrm{true}}$ and dividing by $M_{500,\mathrm{true}}$, while the right column gives the spec mass fractions found using the estimated $r_{500,\mathrm{spec}}$ and $M_{500,\mathrm{spec}}$ for each cluster. For more details on how the spec mass and radii were estimated see Section~\ref{Sec:Mass}. In the left column, the dotted vertical lines show the boundaries of the mass bins defined in Section~\ref{Sec:Analysis}. To show the median trend of the spec mass fractions, spec mass bins were defined such that there were $15\pm1$ clusters in each mass bin at $z=0$ for all the CELR samples:
\begin{itemize}
	\item $M_{500,\mathrm{spec}} > 1.9\times 10^{14}~\mathrm{M}_{\odot}$,
	\item $1.9\times 10^{14} < M_{500,\mathrm{spec}}~[\mathrm{M}_{\odot}] < 6.7\times 10^{14}$,
	\item $M_{500,\mathrm{spec}} > 6.7\times 10^{14}~\mathrm{M}_{\odot}$.
\end{itemize}
The spec mass bins are again shown as vertical dotted lines in the right column. The median trend for each run is then shown in each panel as a solid line, with the 1~$\sigma$ scatter within each mass bin shown by the errorbars. For clarity, we will only comment on the trend of the spec data here and defer a more detailed discussion as to the effect of $M_{500,\mathrm{spec}}$ on the gas/stellar mass fractions to Section~\ref{Sec:HSEMass}. Where individual data points have spec gas or stellar mass fractions which are outside of the plotted region, it is displayed as an arrow at the corresponding $M_{500,\mathrm{spec}}$.

For the gas mass fractions (top row), the grey shaded region shows the combined 1~$\sigma$ scatter of the observed data from \cite{Vikhlinin2006}, \cite{Maughan2008}, \cite{Sun2009} and \cite{Lovisari2015}. All of the clusters have true gas mass fractions (left panel) which are below the universal baryon fraction. Across the whole mass range, the CELR-B clusters (red squares) are in relatively good agreement with the true gas mass fractions of the BAHAMAS project (see Fig.~5, \citealt{McCarthy2017}). The BAHAMAS model was calibrated to reproduce the galaxy stellar mass fraction, but reducing the AGN heating temperature to $\Delta T = 10^{7.8}$~K from $\Delta T = 10^{8}$~K in the previous cosmo-OWLS model \citep{LeBrun2014, McCarthy2017} also resulted in clusters with hot gas mass fractions in good agreement with observations. 

In the lowest mass bin, the true gas mass fractions of the CELR-E/NC clusters (blue circles/green triangles respectively) are higher than those for the CELR-B clusters. As discussed above, between the BAHAMAS and EAGLE projects, the density dependent scale factor used to increase the accretion rate of BHs in low density regions was removed due to the assumed resolution of the simulations. As a result, the CELR-B runs produce median central BH masses across the whole mass range which are more massive by on average a factor of 4. The reduced mass of the BHs in CELR-E and CELR-NC means that there is less energy available for feedback events, so although the total energy released per event is the same (to within $\sim$~25 per cent) for both models, clusters run with the EAGLE model inject less energy and remove less gas at this resolution. Despite the discrepancy in the median BH masses persisting for the higher mass bins there is very little difference in the gas mass fractions between the runs which again highlights how the subgrid physics becomes less important for more massive clusters.

When using estimated $r_{500,\mathrm{spec}}$ and $M_{500,\mathrm{spec}}$, the differences between the runs due to different subgrid models seen in the true gas mass fractions is still present at low mass, with the median values for all of the runs increasing. At higher mass, the median values are also increased and there is a steeper trend in the median spec gas mass fractions than is seen for the true simulation data. This is a result of using estimated spec masses which are likely to be underestimated, especially at high mass (see Section~\ref{Sec:Mass}), leading to increased gas mass fractions. Across the whole mass range, the median of the samples is typically too high compared to the obersvational band, but this is encompassed within the scatter of data.

Comparing the relaxed (filled) and unrelaxed (open) clusters across the three models it can be seen that the relaxed clusters are much more likely to have spec gas mass fractions which lie close to the median line and, therefore, the observational data. Increased scatter when looking at unrelaxed clusters is well documented (e.g. \citealt{Nagai2007a}) as a result of trying to fit models designed for relaxed clusters in hydrostatic equilibrium to objects with large peaks and other features in their profiles.

The bottom row of Fig.~\ref{fig:GasFractions} shows the stellar mass fractions, along with the best fit relationships found by \cite{Giodini2009} and \cite{Budzynski2014}, shown by the grey and black lines respectively. The observed trends are seen to diverge at low masses, which \cite{Budzynski2014} attributes to \cite{Giodini2009} not being able to account for the contribution from intracluster light, found to provide $20~-~40$ per cent of the total stellar mass. Between the true (left) and spec (right) stellar mass fractions, the scatter increases across the whole mass range and the median for the middle and high mass bins increases for all of the models. As with the gas mass fractions, these differences are driven by the use of estimated spec masses. The BAHAMAS project calibrated their model to reproduce the observed galaxy stellar mass function, so the CELR-B clusters have stellar mass fractions which are consistent with the observations (given the uncertainty in the trend at low mass). At high mass, the majority of the CELR-E and CELR-NC clusters have too many stars, especially when using mock spec quantities.

Focusing on just the relaxed clusters (filled points), their spec stellar mass fractions are much more in line with the observations as was seen for the spec gas mass fractions. Unrelaxed clusters again increase the scatter and typically are too stellar rich, especially for the middle and high mass EAGLE based clusters. At low resolution, thermal stellar feedback models are known to be susceptible to radiative losses due to the ratio of heated mass to star particle being too large (e.g. \citealt{Katz1996, Balogh2001}). However, the model of \cite{Dalla2012} works such that the ratio of heated particles to star mass is kept small regardless of the resolution. Decreasing the resolution leads to thermal energy being radiated away before the gas can respond to the increased temperature, due to the sound crossing time-scale across the heated element no longer being short compared to the radiative cooling time-scale in the heated gas. Due to the CELR-E/NC central BHs being smaller than for CELR-B, there may also be overcooling in the gas due to the comparative lack of feedback in the EAGLE runs.

Although recalibrating the EAGLE model could have led to gas and stellar mass fractions for CELR-E and CELR-NC which are more in line with the observations, the relaxed clusters are typically close to or within the scatter of the observed trends which often remove unrelaxed clusters from their analysis. Also, we note that there is little difference between CELR-E and CELR-NC for any of the mass fractions, suggesting that the integrated cluster properties do not sensitively depend on artificial conduction.

\begin{figure}
	\includegraphics[scale=0.5, keepaspectratio=True]{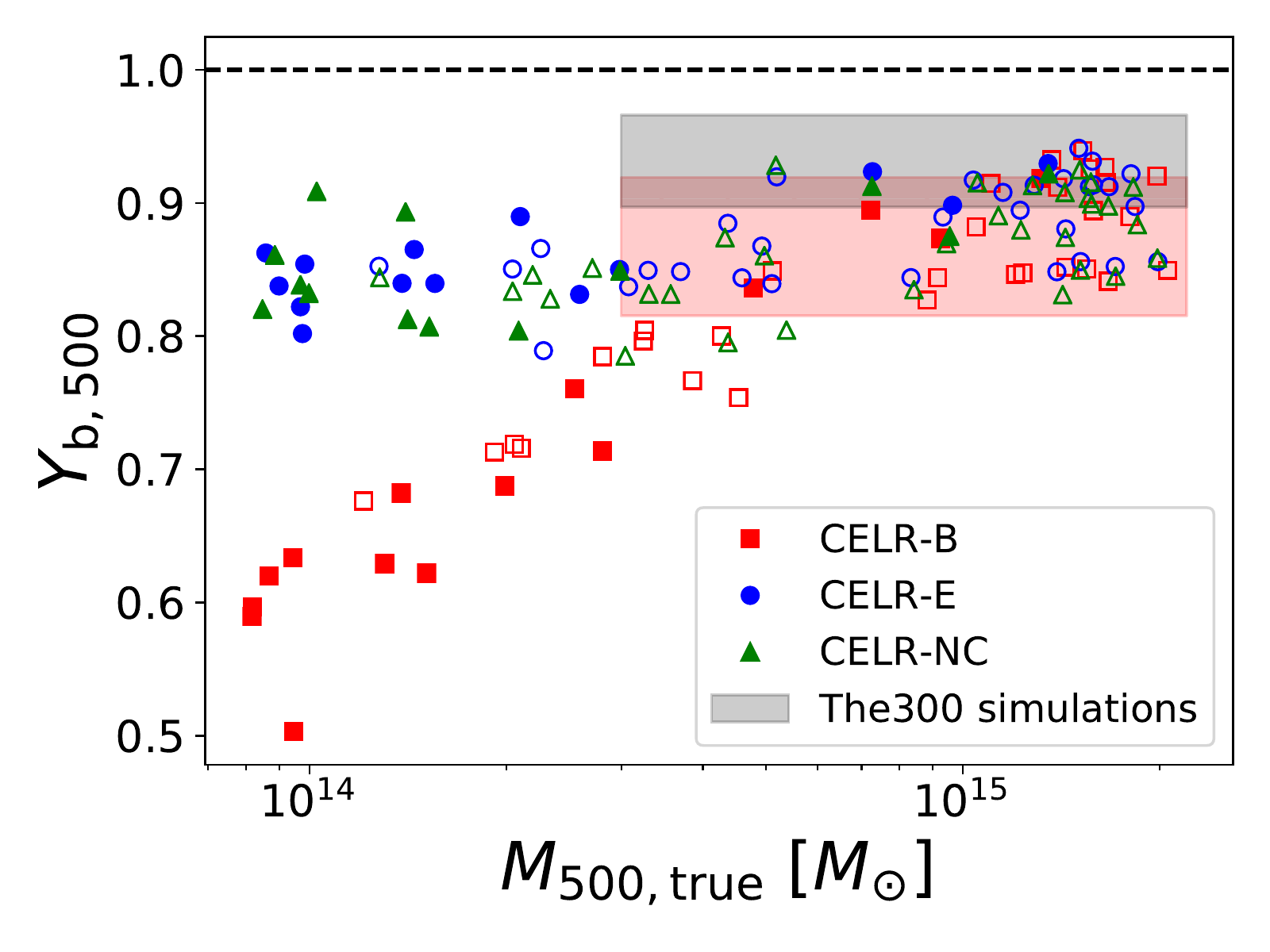}
	\caption{The baryon depletion factor, $Y_{\mathrm{b}}$, for each individual cluster across all three samples. The colour scheme is the same as that in Fig.~\ref{fig:GasFractions}. The black dashed line shows the universal baryon fraction, $\Omega_{\mathrm{b}}/\Omega_{\mathrm{m}} = 0.156$, to which all the values are normalised. The red shaded region shows the median value of $Y_{\mathrm{b}} = 0.852$ for the CELR-B sample and the associated scatter after restricting the mass range of the clusters to $M_{500,\mathrm{true}} > 3.0\times 10^{14}$. For comparison, we also include the results of \citet{Eckert2019} who found $Y_{\mathrm{b}} = 0.938$ when using The300 simulation suite \citep{Cui2018} as the shaded black box.}
	\label{fig:DepletionFactor}
\end{figure}

Fig.~\ref{fig:DepletionFactor} shows the baryon depletion factor, $Y_{\mathrm{b}}$, for all individual clusters in all three runs as a function of mass. The baryon depletion factor is a measure of how close the baryon fraction, $f_{\mathrm{b}}$, of a cluster is to the universal baryon fraction, $\Omega_{\mathrm{b}}/\Omega_{\mathrm{M}} = 0.157$, such that if a cluster had $f_{\mathrm{b}} = \Omega_{\mathrm{b}}/\Omega_{\mathrm{M}}$, then $Y_{\mathrm{b}} = 1$. It can be seen that the distribution of $Y_{\mathrm{b}}$ for the CELR-E/NC clusters is a lot flatter with mass compared to CELR-B, which follows from the combination of the true gas and stellar mass fractions of the clusters in Fig.~\ref{fig:GasFractions}. \cite{Eckert2019} found that for The300 simulation suite \citep{Cui2018}, $Y_{\mathrm{b}}$ was approximately constant for clusters with $M_{500,\mathrm{true}} > 3.0\times 10^{14}$. When restricting the CELR clusters to this mass range, there is still some positive correlation between increasing mass and $Y_{\mathrm{b}}$. Across the whole mass range and the restricted mass range, the $Y_{\mathrm{b}}$ values for CELR-E/NC are always larger than that found for CELR-B, and there is smaller scatter. For comparison, we show the median $Y_{\mathrm{b}}$ for CELR-B and the associated scatter for the restricted mass range as the red shaded area, and the corresponding values for The300 simulation are shown as the black shaded box (taken from \citealt{Eckert2019}). As the BAHAMAS model was calibrated to reproduce observed gas and stellar mass fractions, this suggests that The300 simulations are too gas rich.

\subsection{Density profiles}

\begin{figure*}
	\includegraphics[width=\textwidth,keepaspectratio=True]{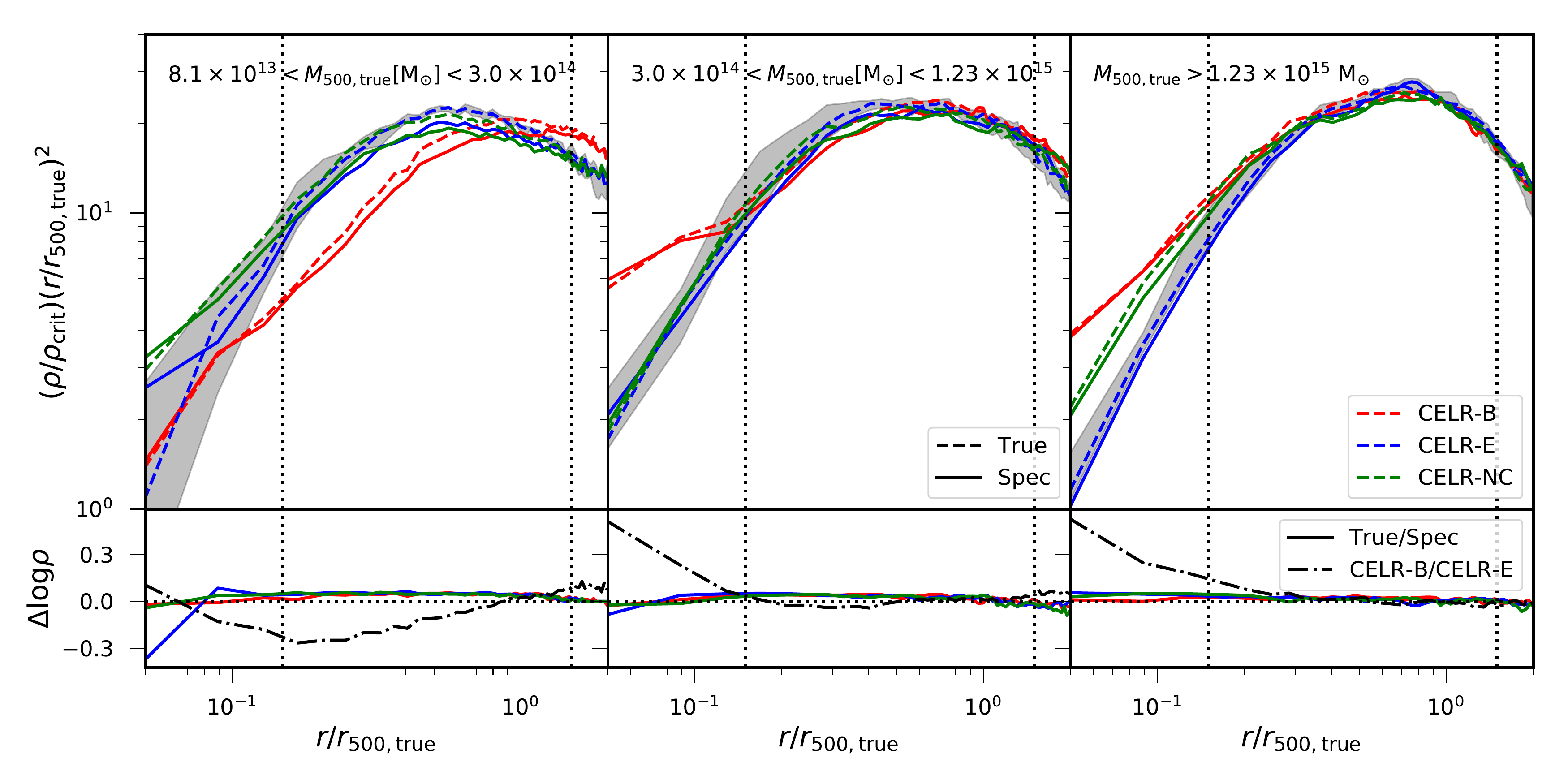}
	\caption{Median density profiles for each mass bin (lowest mass clusters in the left panel, most massive in the right panel) at $z=0$. Red, blue and green lines correspond to CELR-B, CELR-E and CELR-NC respectively. For the top row, the dashed lines show the median true density in each mass bin while the solid lines represent the median profile estimated using mock spectroscopic data. The error on the true CELR-E profile found by bootstrapping the individual profiles is shown by the grey band. The vertical dotted lines show the fitting range 0.15 - 1.5 $r_{500}$ which is used later in the analysis. In the second row, the solid lines show the difference $\Delta\log(\rho) = \log(\rho_{\mathrm{true}}) - \log(\rho_{\mathrm{spec}})$ for all of the runs in the same colour as before. Similarly, the dash-dot black line shows the difference between the CELR-B and CELR-E true profiles.}
	\label{fig:DensityProfile}
\end{figure*}

The median scaled gas density profiles for the low, middle and high mass bins are shown in Fig.~\ref{fig:DensityProfile}. The density values on the $y$-axis are scaled by $(r/r_{500,\mathrm{true}})^{2}$ to reduce the dynamic range. In all three mass bins both the X-ray spectroscopic (solid line) and true density profiles (dashed line) are shown for all three CELR runs. The vertical dotted lines shows the fitting range for the \cite{Vikhlinin2006} density model which is used to determine mass estimates for the clusters in Section~\ref{Sec:Mass} (see Appendix~\ref{Sec:Appendix} for more details). The lower fitting limit of 0.15~$r_{500,\mathrm{true}}$ is greater than the convergence radius of the smallest cluster. The grey band shows the uncertainty on the median true CELR-E profile, found by bootstrapping the clusters within each mass bin. The band encloses the 16th and 84th percentiles of the median profiles of 1000 realisations. The bottom panel shows the value of $\Delta\log(\rho) = \log(\rho_{\mathrm{true}}) - \log(\rho_{\mathrm{spec}})$ for all of the runs in their corresponding colours, and the dash-dot black line is $\Delta\log(\rho)$ between the true CELR-B and CELR-E profiles.

Across the whole mass range, with the exception of the core of the smallest CELR-E clusters, there is a mild offset between the true and spec density profiles. The spec density profiles are lower because the density and metallicity are degenerate with each other when setting the overall normalisation of the spectrum fit. On average we find that the spectroscopic density is marginally lower ($\sim$5 per cent) than the mass-weighted values, and the spectroscopic metallicity is marginally higher ($\sim$5 per cent) using our method. We note that one-to-one line is incorporated within the one sigma scatter envelope of the recovered density and metallicity profiles. Within the core of CELR-E, the spec density increases above the true density profile but is still within the grey errorband.

For the lowest mass clusters the CELR-B median density profile is shifted radially compared to the CELR-E/NC profiles. The peak density of the three samples is relatively constant, but is achieved at a larger radius for the CELR-B clusters. Considering Fig.~\ref{fig:GasFractions} and just the changes in the subgrid physics, it follows that the CELR-B clusters would be less dense in the centre of the cluster as the AGN feedback mechanism used in the BAHAMAS model is more effective at removing gas at this resolution due to their BHs being larger. The displaced gas has been moved to the outskirts of the cluster.

For a non-radiative simulated cluster, as a result of the changing SPH flavour between the runs it would be expected that the core of CELR-B would be more dense than CELR-E. ANARCHY is a pressure-entropy flavour of SPH that includes artificial condution, leading to improved entropy mixing and allowing a single phase gas to be produced which should stop low-entropy gas sinking to the centre of the cluster. However, \cite{Sembolini2016b} found that when simulating a single cluster using 10 cosmological models with different hydrodynamics and radiative physics, changes to the subgrid models had a bigger impact on the final cluster than differences in the SPH flavour. Within the core of the low mass clusters, the CELR-NC profile is the most dense, lying just outside of the error for CELR-E (the grey band) right at the centre of the cluster. The true density profile for CELR-E also diverges from the spec density in the very centre of the cluster such that the true CELR-E profile is the least dense. This shows that although the subgrid physics has a larger impact on the final cluster as CELR-B is the least dense out to $r_{500,\mathrm{true}}$ with the exception of the very centre of the cluster, the introduction of both the pressure-entropy formalism and artificial conduction does lead to CELR-E being less dense than the two models without artificial conduction in the very centre due to the lack of low entropy gas sinking to the cluster core.

As the mass of the cluster increases, the CELR-B profiles become more comparable to the CELR-E and CELR-NC profiles between $0.15~-~1.5~r_{500}$. For the middle and high mass bins, gas in the CELR-B clusters becomes considerably more dense than for the other two samples as the radius decreases from $0.15~r_{500}$ towards the cluster centre. In the highest mass clusters there is a bigger difference between the CELR-E and CELR-NC profiles than in the other mass bins; the clusters without artificial conduction are more dense and lie outside of the error on CELR-E. Thus, the impact of the artificial conduction implemented in ANARCHY is clearly visible in the cores of high mass clusters.

\subsection{Temperature profiles}\label{Sec:Temperature}

\begin{figure*}
	\includegraphics[width=\textwidth,keepaspectratio=True]{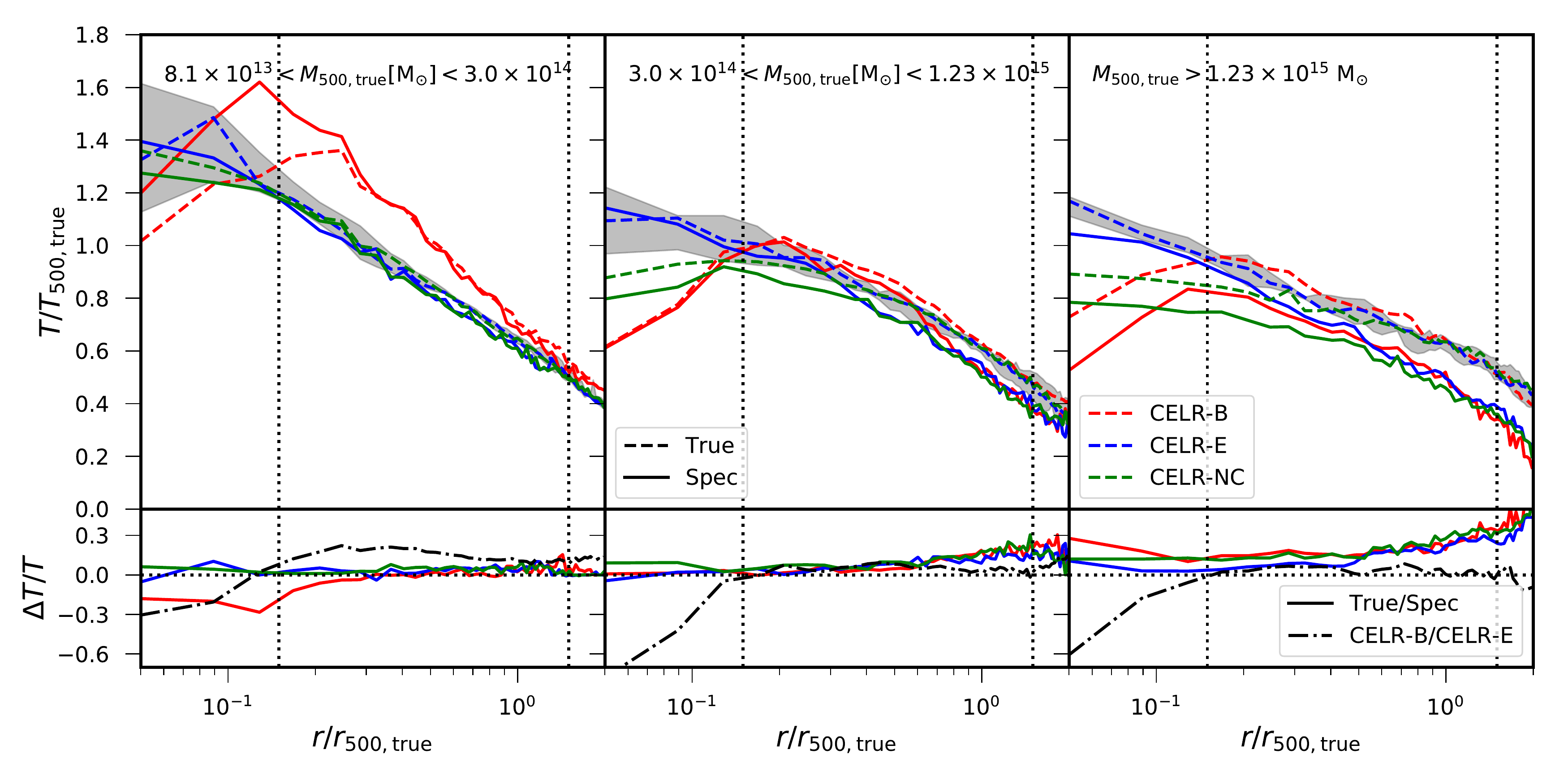}
	\caption{The median temperature profiles following the same layout, linestyle and colour scheme as Fig.~\ref{fig:DensityProfile}. Individual clusters are normalised by their $T_{500}$ before the median value in each mass bin is taken. The lower panels show the fractional difference, $\Delta T/T = (T_{\mathrm{true}}-T_{\mathrm{spec}})/T_{\mathrm{true}}$, between the true and spec profiles for each run in the same colour as the temperature profiles while the dash-dot black line gives the fractional difference between the true profiles of CELR-B and CELR-E, $\Delta T/T = (T_{\mathrm{CELR-B}}-T_{\mathrm{CELR-E}})/T_{\mathrm{CELR-E}}$.}
	\label{fig:TemperatureProfile}
\end{figure*}

Fig.~\ref{fig:TemperatureProfile} shows the median scaled temperature profiles in the same layout as Fig.~\ref{fig:DensityProfile}. At all masses there is a noticeable difference in the shape of the profiles between the different runs. The CELR-B profile peaks between $0.1-0.4~r_{500,\mathrm{true}}$ regardless of mass, whereas the CELR-E/NC profiles do not peak within our radial range. The CELR-E mass-weighted temperature continues to increase into the core while the CELR-NC profile appears to flatten to a lower temperature. By turning off the artificial conduction, the gas is no longer mixing to the same extent that it does in the CELR-E clusters and is instead becoming more like the CELR-B profile where low entropy gas sinks to the core of the halo and lowers the median temperature.

For the smallest clusters, the CELR-B profile also has a steeper decline in temperature at larger radii. This follows from the density profiles (Fig.~\ref{fig:DensityProfile}) which shows that the CELR-B sample is less dense than the other samples within $\sim r_{500,\mathrm{true}}$, for both the true and spec profiles, and can also be seen in the top row of Fig.~\ref{fig:MassMap}. Due to their reduced density, the CELR-B clusters must be hotter in order to maintain pressure support against gravitational collapse. Unlike the density profiles, the temperature of the CELR-B clusters is greater than for CELR-E/NC at all radii outside of the core ($> 0.15~r_{500}$) for the smallest clusters. As the cluster mass increases, CELR-B becomes more in line with the CELR-E temperature profile outside of the core.

As the cluster mass increases, the true temperature profiles become increasingly hotter than their corresponding spec profiles. When fitting to the spectrum, a single-temperature model is used despite the abundance of multi-temperature gas, biasing the spectroscopic temperature low with respect to the mass-weighted temperature (e.g. \citealt{Mazzotta2004}). The discrepancy is known to get worse as the spread of temperatures in the clusters increases towards the cluster outskirts \citep{Mazzotta2004}. This is expected in more massive clusters which are dynamically younger so typically have higher rates of gas accretion and more substructures in their outskirts. Within the core of the clusters, the CELR-E runs have spec and true profiles which are most similar across the whole mass range, suggesting that the inclusion of artifical conduction helps to reduce the bias between the profiles at small radii.

\subsection{Thermal pressure profiles}

\begin{figure*}
	\includegraphics[width=\textwidth,keepaspectratio=True]{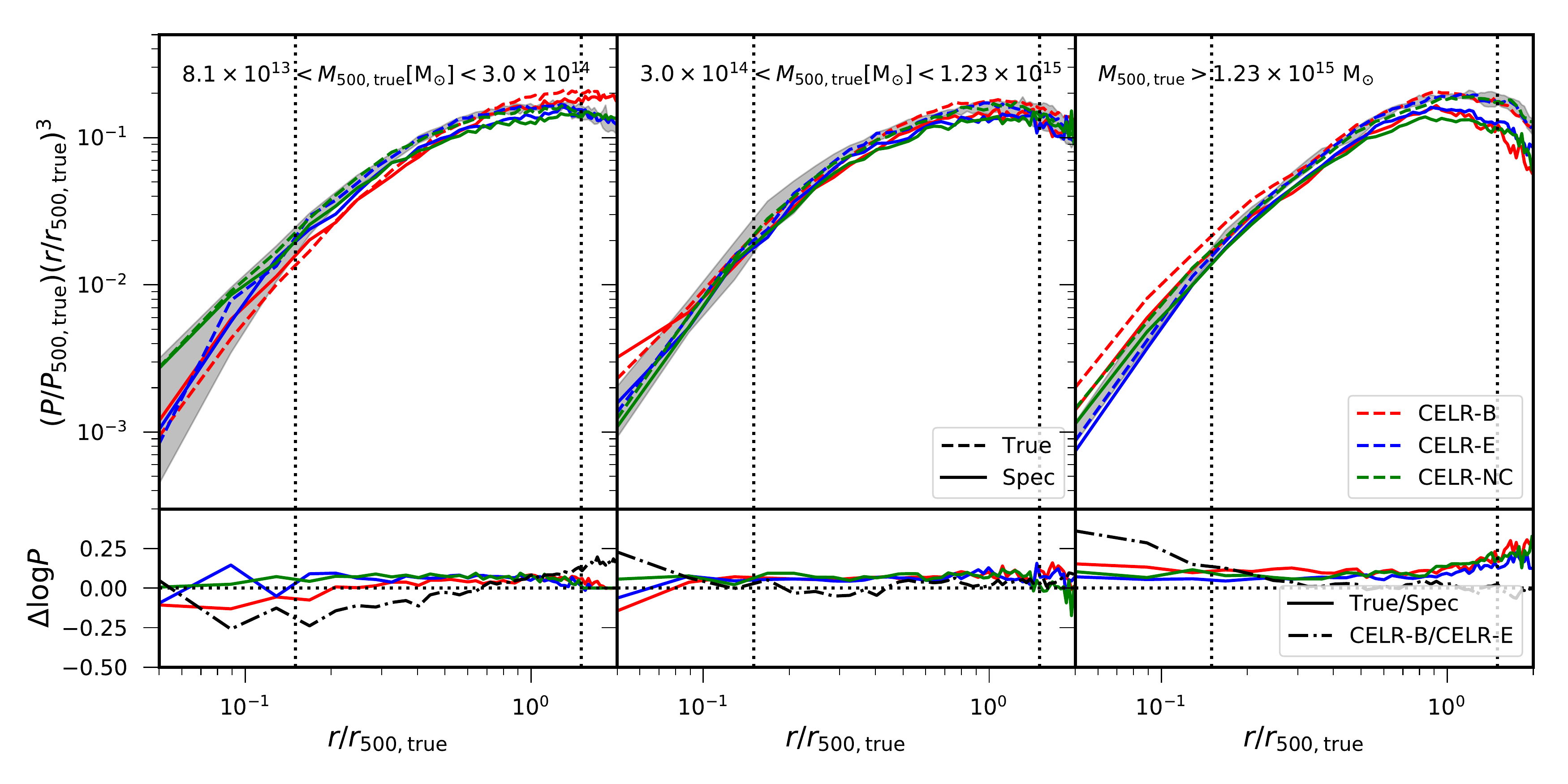}
	\caption{The median pressure profiles following the same layout, linestyle and colour scheme as Fig.~\ref{fig:DensityProfile}. Within each mass bin, the individual profiles are normalised by $P_{500}$ before the median profile is determined. The dynamic range of the data is reduced by scaling the profiles by $(r/r_{500,\mathrm{true}})^{3}$. In the bottom panels, the solid lines represent $\Delta\log P = \log P_{\mathrm{true}} - \log P_{\mathrm{spec}}$ for each run in the same colour as for the pressure profiles. The dash-dot black line is the difference between the true profiles for CELR-B and CELR-E.}
	\label{fig:PressureProfile}
\end{figure*}

The hot gas pressure profiles were calculated by directly combining the density and temperature profiles according to
\begin{equation}\label{eq:Pth}
P_{\mathrm{th}}(r) = \left(\dfrac{k_{\mathrm{B}}}{\mu m_{\mathrm{p}}}\right) \rho(r)T(r),
\end{equation}
assuming $\mu=0.59$, the mean molecular weight of the gas. Fig.~\ref{fig:PressureProfile} shows the median scaled pressure profiles for all of the runs to enable direct comparison of the different samples. The profiles are scaled by $(r/r_{500,\mathrm{true}})^{3}$ to reduce the dynamic range of the data.

From the residuals, there is an offset between the true CELR-B and CELR-E profiles for the smallest clusters. For $0.15~<~r [r_{500}]~<~0.8$, the CELR-B sample has a lower pressure compared to CELR-E by up to 40 per cent, but is higher by up to 25 per cent at larger radii. A similar result is seen in the density profiles (Fig.~\ref{fig:DensityProfile}). However, in the temperature profiles (Fig.~\ref{fig:TemperatureProfile}), CELR-B is marginally hotter than CELR-E for $r > 0.15~r_{500}$, so the pressure differences are predominatly driven by the density.

Again, as the mass of the clusters increases, the profiles become more similar. For the middle mass bin, CELR-B also has a higher pressure compared to CELR-E at large radii, but this occurs at a smaller radius, $r\sim 0.4~r_{500}$. In the highest mass bin, the CELR-B and CELR-E pressure profiles are indistinguishable at $r > 0.4~r_{500}$. The difference between the true and spec profiles increases with mass, which is driven by the true (mass-weighted) temperature increasing above the spectroscopic temperature at higher mass.

\subsection{Non-thermal pressure profiles}

\begin{figure*}
	\includegraphics[width=\textwidth,keepaspectratio=True]{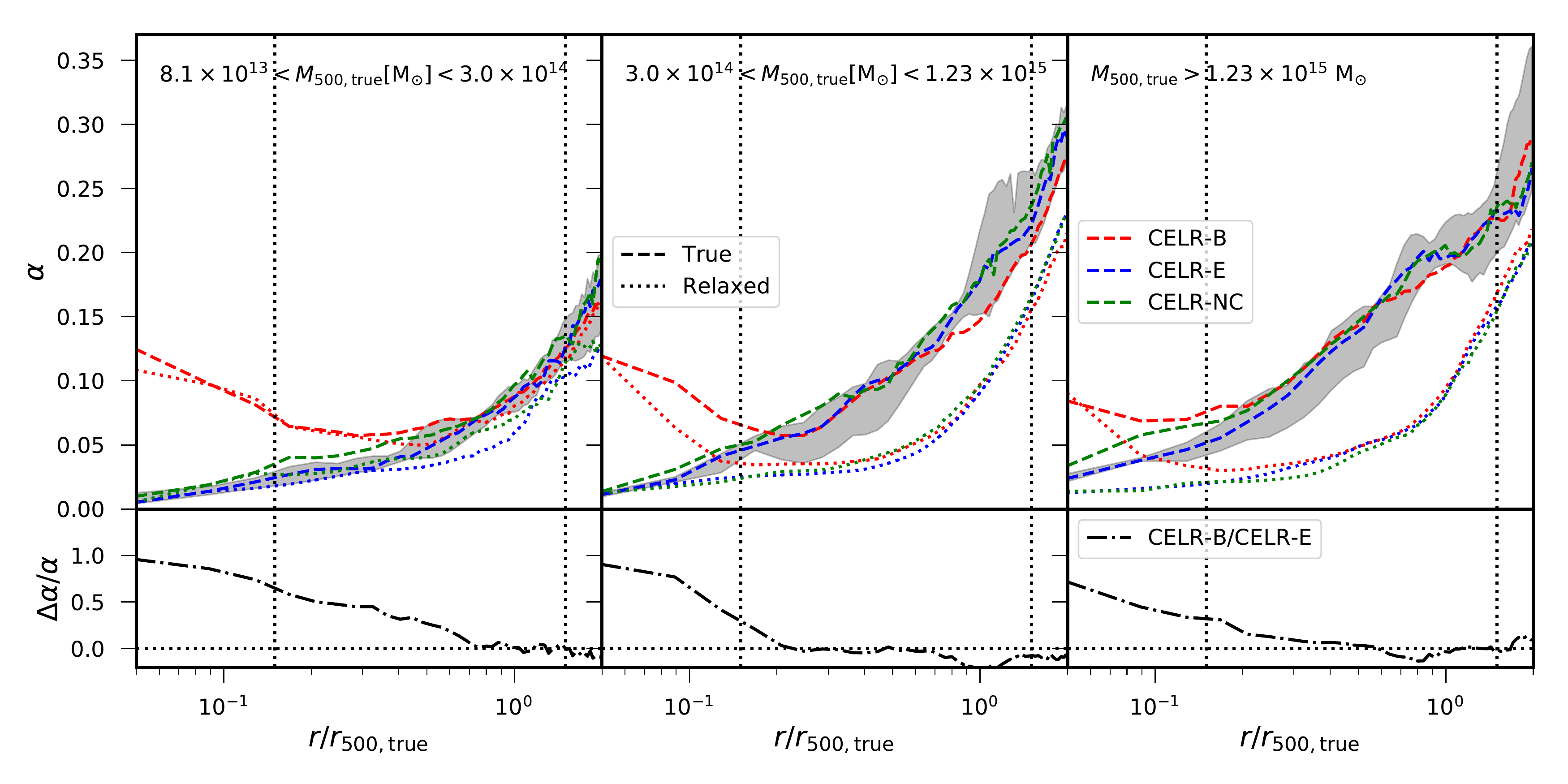}
	\caption{The fractional contribution of non-thermal pressure to the total pressure following the same layout, linestyle and colour scheme as Fig.~\ref{fig:DensityProfile}, although only the true profiles are presented here. We also present the relaxed cluster subset as a dotted line in each mass bin. The dash-dot black line in each lower panel gives the fractional difference between the CELR-B and CELR-E profiles, $\Delta\alpha/\alpha = (\alpha_{\mathrm{CELR-B}}-\alpha_{\mathrm{CELR-E}})/\alpha_{\mathrm{CELR-E}}$.}
	\label{fig:PnthProfile}
\end{figure*}

Non-thermal pressure, in the form of bulk motions and turbulence in the gas (neglecting other sources such as magnetic fields and cosmic rays), is expected to contribute as much as 30 per cent of the overall pressure at $r_{500}$ (e.g \citealt{Nelson2014a}). The non-thermal component of the total pressure can be characterised as
\begin{equation}
P_{\mathrm{nth}} = \alpha(r)P_{\mathrm{tot}},
\end{equation}
assuming $P_{\mathrm{tot}} = P_{\mathrm{th}} + P_{\mathrm{nth}}$.
From simulations, the non-thermal pressure can be estimated using
\begin{equation}\label{eq:Pnth}
P_{\mathrm{nth}} = \rho\sigma^{2},
\end{equation}
where $\sigma^{2} = \sum_{j}\sigma_{j}^{2}$ is the three-dimensional ($j \in \{x,y,z\}$) squared velocity dispersion of the gas measured in shells around the centre of the cluster and
\begin{equation}
\sigma_{j}^{2} = \dfrac{\sum_{i} m_{i}(v_{i,j}-\bar{v}_{j})^{2}}{\sum_{i}m_{i}},
\end{equation}
where $m_{i}$ is the mass of the $i$th particle in the shell and $\bar{v}$ is the mean velocity of the shell.

Fig.~\ref{fig:PnthProfile} shows the ratio of non-thermal to total pressure, $\alpha(r)$, in each mass bin. Only the true profiles are shown as the spec model for $P_{\mathrm{nth}}$ would require us to model $\sigma$ from X-ray line widths which we have not attempted for the current work. We also present $\alpha$ for the relaxed cluster subset in each mass bin (dotted line) for all the runs.

Regardless of the cluster mass or dynamical state, the $\alpha$ profile for CELR-E/NC increases across the whole radial range, whereas for CELR-B it decreases or remains constant until $r \simeq 0.3~r_{500,\mathrm{true}}$ and then increases out to at least 2~$r_{500,\mathrm{true}}$. Comparing the thermal and non-thermal pressure profiles for individual clusters, the overall shape of $\alpha$ is driven by $P_{\mathrm{nth}}$, suggesting that the increase in non-thermal pressure fraction towards the cluster centre for CELR-B is due to AGN feedback creating larger bulk motions in the core. As the cluster mass increases, the median CELR-B profile flattens in the core and becomes more comparable to the CELR-E/NC clusters due to the thermal pressure in the core of CELR-B being greater than in CELR-E/NC (see Fig.~\ref{fig:PressureProfile}). This behaviour is not seen in the relaxed subset of CELR-B which instead continues to increase towards the centre in all mass bins.

Looking at the shape of the $\alpha$ profiles, for $r > 0.15~r_{500,\mathrm{true}}$ there is no difference due to the dynamical state of the clusters in the smallest mass bin for any of the runs as most clusters are defined to be relaxed at this mass. As cluster mass increases, the median $\alpha$ profile for all clusters in each bin increases approximately linearly with radius. For the relaxed clusters, of which there are considerably fewer at high mass, $\alpha$ remains relatively constant at around 4 per cent until 0.7~$r_{500,\mathrm{true}}$ at which point $\alpha$ increases rapidly out to at least 2~$r_{500,\mathrm{true}}$. This upturn at large radii can be attributed to accretion of gas in the cluster outskirts causing non-thermal motions (e.g. \citealt{Nelson2014b,Shi2014}). At all radii, the fractional contribution of $P_{\mathrm{nth}}$ to total pressure is less for the relaxed clusters for all three runs, as expected.

The maximum value of $\alpha$ increases with the mass of the clusters from around 10 per cent at $r_{500}$ in the lowest mass bin to $\sim$~25 per cent for the highest mass bin. \citeauthor{Shi2016} (\citeyear{Shi2014, Shi2016}) found that the non-thermal pressure correlates with this mass history of a cluster as the dominant source of bulk motions is major mergers. Since the most massive clusters have formed more recently compared to smaller clusters, they would be expected to have a higher non-thermal pressure contribution. This can also be seen in the fraction of unrelaxed clusters which increase from around 30 per cent in the lowest mass bin to $> 90$ per cent in the highest mass bin for all of the runs. Across the whole mass range, the CELR-E clusters have the lowest contribution from non-thermal pressure in the core, becoming more significant for the most massive clusters. This highlights how artificial conduction has the most noticeable effect in the core of massive clusters and is able to smooth over discontinuities in the gas. For the relaxed subset, there is very little difference between the CELR-E and CELR-NC $\alpha$ profiles in the centre of the more massive clusters, suggesting that artificial conduction is most important in objects undergoing mergers.

Observationally, it is very difficult to measure the non-thermal pressure fraction of clusters. Data from the \textit{Hitomi} spacecraft allowed the ratio of turbulent-to-thermal pressure to be measured in the Perseus cluster as 4 per cent \citep{Hitomi2016}. However, this measurement was limited to within the cluster core ($r<60$~kpc) where the ratio is expected to be small (although this was not seen for the CELR-B clusters). Instead, \cite{Ghirardini2018} suggest a method for estimating the non-thermal pressure using the total baryon fraction in massive clusters. This method is based on the principle that the cluster baryon fraction is expected to be close to the universal baryon fraction as clusters originate from overdensities in the primordial universe, such that
\begin{equation}
f_{\mathrm{gas,univ}} = Y_{\mathrm{b}}\dfrac{\Omega_{\mathrm{b}}}{\Omega_{\mathrm{m}}} - f_{*},
\end{equation} 
where $f_{\mathrm{gas,univ}}$ is the universal cluster gas mas fraction, $Y_{\mathrm{b}}$ is the baryon depletion factor which is the ratio of the baryon fraction to the universal value, $\Omega_{\mathrm{b}}/\Omega_{\mathrm{r}}$, and $f_{*}$ is the stellar mass fraction. A measure of the non-thermal pressure in a cluster can then be found by comparing the measured $f_{\mathrm{gas}}(r)$ of a cluster to $f_{\mathrm{gas,univ}}$ (see Eq.~17, \citealt{Ghirardini2018}).

\cite{Eckert2019} used this method to try and place a limit on the non-thermal pressure contribution in the X-COP cluster sample \citep{Eckert2017}. To calculate $f_{\mathrm{gas,univ}}$, The300 simulations \citep{Cui2018} were used to find the baryon depletion factor, $Y_{\mathrm{b}} = 0.938^{+0.028}_{-0.041}$. A value of $f_{*} = 0.015\pm 0.005$ was found by combining multiple obersavtional datasets (see Fig.~3, \citealt{Eckert2019} and references therein). Combining these with the \cite{PlanckXIII2016} value $\Omega_{\mathrm{b}}/\Omega_{\mathrm{m}} = 0.156\pm 0.003$, $f_{\mathrm{gas,univ}} = 0.131\pm 0.009$. Comparing the values of $f_{\mathrm{gas}}$ to the universal gas fraction for each cluster, \cite{Eckert2019} found that the X-COP sample was on average 7 per cent more gas rich than $f_{\mathrm{gas,univ}}$, and when solving for $\alpha$ they were able to place an upper limit of 13 per cent on the mean level of non-thermal pressure in the Planck cluster population. This limit is lower than the levels of $P_{\mathrm{nth}}$ found here and in other works (e.g. \citealt{Nelson2014b}). However, this limit is dependent on the value of $f_{\mathrm{gas,univ}}$.

When using the CELR-B cluster sample, which uses the BAHAMAS model that is calibrated to reproduce observed gas mass fractions, we determined $Y_{\mathrm{b}} = 0.852^{+0.046}_{-0.037}$ across the same mass range as used for The300 simulations, where the errors give the 1$\sigma$ scatter (see Fig.~\ref{fig:DepletionFactor}). Using the same values of $\Omega_{\mathrm{b}}/\Omega_{\mathrm{m}}$ and $f_{*}$, we find $f_{\mathrm{gas,univ,B}} = 0.118^{+0.006}_{-0.001}$ where the errors have been found by bootstrapping the data. This is considerably smaller than that found by \cite{Eckert2019}, leading to the X-COP clusters being on average around 20 per cent more gas rich than the universal gas fraction found using CELR-B. We leave a detailed calculation on the effect of changing $f_{\mathrm{gas,univ}}$ on $\alpha$ to future work, but here we note that our smaller value of $f_{\mathrm{gas,univ}}$ will lead to an increase in the limit of non-thermal pressure in the X-COP clusters, assuming the shape of the non-thermal profile does not change.

\section{Hydrostatic mass bias}\label{Sec:Mass}

In Section~\ref{Sec:GasProperties}, we introduced the different hot gas profiles necessary to estimate the hydrostatic mass profile of a cluster. Comparing the results from the different runs, it was found that at low mass ($8\times 10^{13} < M_{500,\mathrm{true}}~[\mathrm{M}_{\odot}] < 3\times 10^{14}$) the impact of the subgrid physics was greater than the effects of changing the SPH flavour (as was also reported by \citealt{Sembolini2016b}). However, when moving to the most massive clusters ($M_{500,\mathrm{true}} \gtrsim 10^{15}$ \msol), the effect of changing the SPH flavour can be seen in the cluster cores, whereas there is less of an effect from changing the subgrid physics as the cluster potentials are too deep. Outside of $0.15~r_{500,\mathrm{true}}$ there is little difference between the runs for any of the profiles. In this section, we will discuss hydrostatic mass estimates using different combinations of the hot gas profiles, and check whether, as out results suggest, the mass estimates (when measured at $r_{500}$) are insensitive to variations in the subgrid physics and SPH method.

The clusters are modelled as being spherically symmetric and in hydrostatic equilibrium, thus satisfying
\begin{equation}\label{eq:HSE}
\dfrac{\mathrm{d}P}{\mathrm{d}r} = -\rho\dfrac{GM_{\mathrm{HSE}}(<r)}{r^2},
\end{equation}
where $M_{\mathrm{HSE}}$ is the hydrostatic mass of the cluster (equal to the true mass if the cluster is spherical and perfectly hydrostatic). In practice, solving this equation for $M_{\mathrm{HSE}}$ only leads to an estimate of the mass of a cluster. This can then be used to define the hydrostatic mass bias;
\begin{equation}
b = 1 - M_{\mathrm{HSE}}/M_{\mathrm{true}},
\end{equation}
which quantifies the cluster's departure from hydrostatic equilibrium. However, we note that $b$ could also encompass other systematic offsets between the cluster masses such as e.g. instrument calibration as well as the effects of non-thermal pressure and temperature inhomogeneties. To address the latter, we also apply a correction to the cluster mass estimates to account for the non-thermal pressure (Fig.~\ref{fig:PnthProfile}) similar to e.g. \cite{Nelson2014b, Martizzi2016} and \cite{Shi2016}.

\subsection{Summary of recent studies}

Recently, \cite{Henson2017} found that the hydrostatic mass bias of the combined MACSIS \citep{Barnes2017a} and BAHAMAS \citep{McCarthy2017} samples of simulated clusters was mass dependent (see also \citealt{Barnes2017b}). This is in contrast to nearly all previous studies on both observed and simulated clusters which have found that the bias is constant but disagree on the extent of the bias. However, these studies were limited to a smaller mass range with few massive clusters ($M_{500} > 10^{15}$~M$_{\odot}$) in particular.

Focusing on observed clusters, the mass bias is often defined as $\bx = 1 - M_{\mathrm{X}}/M_{\mathrm{WL}}$ where $M_{\mathrm{X}}$ is the hydrostatic mass obtained from X-ray data and $M_{\mathrm{WL}}$ is the weak lensing determined mass. Although it has been shown that $M_{\mathrm{WL}}$ can be biased low by $\sim~5-10$~per cent due to projection effects \citep{Becker2011, Bahe2012, Rasia2012, Henson2017}, weak lensing masses are still assumed to better reflect the true mass. Some groups have found that the X-ray masses of clusters are essentially unbiased compared to $M_{\mathrm{WL}}$ with \cite{Israel2014} and \cite{Smith2016} reporting $\bx = 0.08$ and $\bx = 0.05$ respectively. Conversely, \cite{vonderLinden2014} and \cite{Hoekstra2015} found that $\bx = 0.30$ and $\bx = 0.24$ respectively. All of these analyses were carried out on clusters spanning an overlapping mass range, so the difference cannot obviously be attributed to mass. Interestingly, \cite{vonderLinden2014} found possible evidence for a mass dependence of $\bx$ but explain this away as a result \textit{XMM}-Newton measurements of clusters being $\sim$ 20 per cent lower than those from \textit{Chandra} for massive clusters with $\kb T > 5~\mathrm{keV}$ (\citeauthor{Mahdavi2013} \citeyear{Mahdavi2013}, \citeauthor{Schellenberger2015} \citeyear{Schellenberger2015}). This can lead to mass biases of around 30 per cent for massive clusters, whereas smaller clusters should not be affected.

At low mass ($\sim10^{14}$~\msol), simulations generally find more consistent results with a mass bias of $b \sim 0.2$ when using the simulation data (e.g. \citealt{Kay2012,Biffi2016}) and when attempting to simulate mock observations (e.g. \citealt{Nagai2007a,Rasia2012,LeBrun2014,Barnes2017b,Henson2017}). However, due to the expense of simulating clusters these analyses were focused on samples with no, or very few, high mass objects ($M_{500,\mathrm{true}} > 10^{15}~\mathrm{M}_{\odot}$) so were not able to characterise the mass dependence of the hydrostatic mass bias. \cite{Henson2017} found that, at worst, $b$ increased from 0.2 to 0.4 for clusters with mass increasing from $10^{14} h^{-1}~\mathrm{M}_{\odot}$ to $10^{15} h^{-1}~\mathrm{M}_{\odot}$, similar to the bias of $\sim 0.25-0.35$ found by \cite{Rasia2012}. The bias was found to change when using different combination of spec and true density and temperature profiles, with the largest bias introduced from the X-ray temperature profile. Recently, \cite{Medezinski2018} presented the SZ masses of five \textit{Planck} clusters which suggest that bias could be a function of mass, although the conclusions are limited due to the small sample size.

\subsection{Hydrostatic mass estimates}\label{Sec:HSEMass}

\begin{figure*}
	\includegraphics[width=\textwidth,keepaspectratio=True]{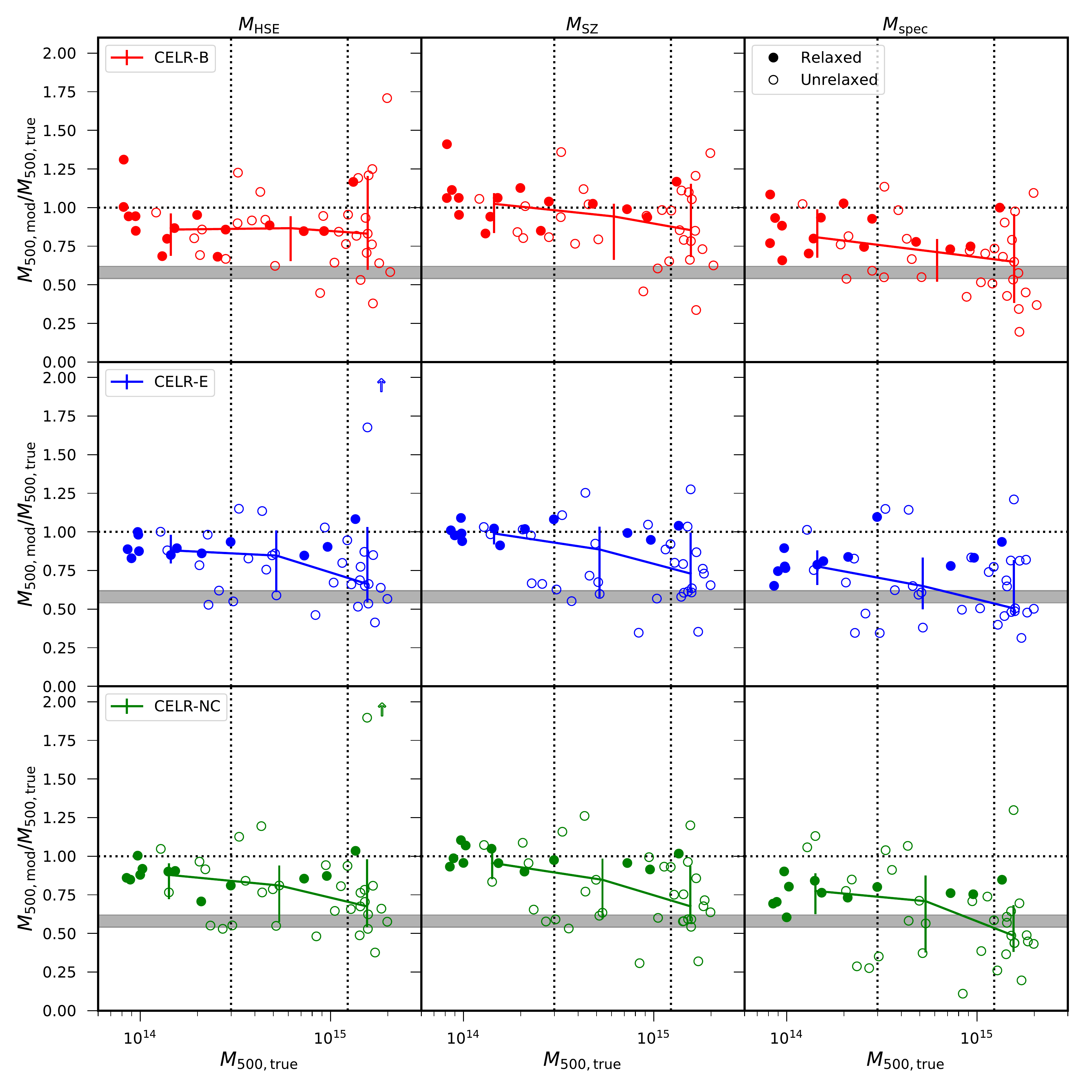}
	\caption{A measure of the offset between the estimated hydrostatic mass compared to the true mass of a cluster plotted against the true mass of the cluster. The top row shows the CELR-B sample, the middle is CELR-E and the bottom row is CELR-NC. From left to right the mass is determined by fitting models to the true density and temperature profiles, to the true pressure profile and specy density profile to try to recreate an SZ mass, and finally to the spec density and temperature profiles. The lines show the median mass bias of the clusters in each mass bin, with the error bars showing the $\sim$1~$\sigma$ scatter. Where an individual cluster has $M_{500,\mathrm{mod}}/M_{500,\mathrm{true}} > 2$, it is displayed by an arrow.}
	\label{fig:Bias}
\end{figure*}

We consider three different ways of estimating the mass of a cluster: HSE, SZ and spec, represented by $M_{\mathrm{HSE}}$, $M_{\mathrm{SZ}}$ and $M_{\mathrm{spec}}$ respectively. All of the mass estimates are found using different combinations of the hot gas profiles (presented in Section~\ref{Sec:GasProperties}). Both $M_{\mathrm{HSE}}$ and $M_{\mathrm{spec}}$ can be calculated by solving Eq.~\ref{eq:HSE} using the definition of pressure as a combination of temperature and density (Eq.~\ref{eq:Pth})
\begin{equation}\label{eq:massTX}
M_{\mathrm{mod}} = -\dfrac{k_{\mathrm{B}}Tr}{G\mu m_{\mathrm{H}}}\left[\dfrac{\mathrm{d}\ln\rho_{\mathrm{gas}}}{\mathrm{d}\ln r} + \dfrac{\mathrm{d}\ln T}{\mathrm{d}\ln r}\right],
\end{equation}
where $M_{\mathrm{mod}} = M_{\mathrm{HSE}}$ uses the true simulation data for the temperature and density profiles, and $M_{\mathrm{mod}} = M_{\mathrm{spec}}$ uses the spec data profiles. The SZ mass of a cluster is found by solving
\begin{equation}\label{eq:massSZ}
	M_{\mathrm{SZ}} = -\dfrac{rP_{\mathrm{th}}}{G\rho_{\mathrm{gas}}}\dfrac{\mathrm{d}\ln P_{\mathrm{th}}}{\mathrm{d}\ln r}.
\end{equation}
We refer to this as the SZ mass as it corresponds to the mass found by combining the true (i.e. mass-weighted) pressure profile (which should be recoverable from SZ data) and the spec density profile. The main difference between SZ and spec masses is that the latter requires X-ray spectroscopy to measure the $T$ profile.

The mass of each cluster was found by first fitting models to the density, temperature and pressure profiles. In this analysis, the true pressure profiles are modelled as a generalised NFW model \citep{Arnaud2010, PlanckV2013}. For both the density and temperature profiles (true and spec), the respective \cite{Vikhlinin2006} models are used (for more information on the model fitting see Appendix~\ref{Sec:Appendix}). An analytic expression for $M_{\mathrm{mod}}(<r)$ was then found by combining the different models according to Eq.'s~\ref{eq:massTX} and \ref{eq:massSZ}. The values of $M_{500,\mathrm{mod}}$ and $r_{500,\mathrm{mod}}$ were then found by interpolation of $M_{\mathrm{mod}}(<r)$.

Fig.~\ref{fig:Bias} shows the individual values of $M_{500,\mathrm{mod}}/M_{500,\mathrm{true}}$ $(=~1-b)$ for all of the CELR samples; the top panel shows CELR-B, the middle is CELR-E and the bottom is CELR-NC. Relaxed clusters are represented with filled circles while unrelaxed clusters are open. If any clusters have a bias value outside the range $0 < 1-b < 2$ an arrow is plotted at the corresponding $M_{500,\mathrm{true}}$ value. The first column corresponds to $M_{500,\mathrm{mod}}~=~M_{\mathrm{HSE}}$, the middle is $M_{500,\mathrm{mod}}~=~M_{\mathrm{SZ}}$ and the final column is $M_{500,\mathrm{mod}}~=~M_{\mathrm{spec}}$. The coloured line shows the trend of the median value of bias in each mass bin, with the errorbars highlighting the 16th and 84th percentile scatter. The grey band in all of the subplots corresponds to the value of $1 - b  = 0.58\pm 0.04$ needed to reconcile the tension in the cosmological parameter estimates between the CMB and SZ analyses \citep{PlanckXXIV2016}.

Focusing on $M_{\mathrm{HSE}}$, the CELR-B sample is consistent with a mass bias of $\sim$~10 - 20 per cent ($1-b$ = 0.9 - 0.8) across the whole mass range. This is similar to the result of \cite{Henson2017} with the combined MACSIS and BAHAMAS sample when also using true gas and density profiles. For CELR-E and CELR-NC, the median $M_{\mathrm{HSE}}$ is consistent with $1 - b~\sim~0.9$ for clusters with $M < 1.23\times 10^{15}$~\msol~but dips for the highest mass clusters as a result of the scatter with the overall distribution of cluster mass estimates being similar.

When using the true mass-weighted pressure and spectroscopic density to obtain $M_{\mathrm{SZ}}$, the median $1-b$ for the smallest mass clusters ($8.1\times 10^{13} < M_{\mathrm{500,true}}~$[\msol]$~< 3.0\times 10^{14}$) is consistent with no bias for all of the samples. As mass increases, $1-b$ decreases and the scatter in the measurements increases. Comparing $M_{\mathrm{SZ}}$ and $M_{\mathrm{HSE}}$, the two methods produce very consistent mass estimates for the most massive clusters with similar median values and level of scatter, whereas the SZ method is more accurate at low mass. This is driven by the spec density profile and not the use of different profile models as using the true density in the SZ method produces very similar results to $M_{\mathrm{HSE}}$.

There is more evidence for a mass dependence of $1 - b$ when using the density and temperature profiles derived from mock 3D spectra and trying to get closer to the analysis done by X-ray observations, although we do not consider the effects of projection in this work. Compared to the SZ mass estimates, there is a similar slope in the median line but a decrease of $\sim$~20 per cent in the normalisation of each mass bin for all of the simulations. This increase in the bias is attributed to the spec temperature profile (e.g. \citealt{Rasia2012, Biffi2016, Henson2017}) and the difficulties in fitting a single temperature fit to data which spans a wide range of temperatures, particularly in high mass clusters. For the most massive clusters, the median bias is in the regime necessary to reconcile CMB and SZ cluster counts with $1 - b\sim 0.4-0.5$. For all of the runs the $1~\sigma$ scatter reaches above $1 - b = 0.62$, but there are still a number of clusters below the lower bound of $1 - b = 0.54$ from \cite{PlanckXXIV2016}.

Changing between \textsc{gadget}-SPH and ANARCHY (CELR-B to CELR-E), it was thought that the mass bias might increase due to improved mixing of the gas phases leading to less substructure within the clusters. \cite{Mathiesen1999b, Nagai2011, Roncarelli2013} found that the uncertainty introduced by the presence of accreting substructures was around 10 per cent of the true gas mass at $r_{200}$ which propogates into the mass bias. However, comparing all of the different models for estimating the cluster mass for the three different simulations, it can be seen that there is no reduction in the mass bias as a result of changing to ANARCHY. The effect of artificial conduction is most evident in the cores of clusters (see Section~\ref{Sec:GasProperties}) and having artificial conduction turned on did not significantly change the distribution or amount of substructure within the clusters.

Across all of the runs and for all of the methods used to estimate the mass of a cluster, there is a small number of objects which scatter to very high or very low bias values. Looking at the individual profiles of these clusters, they all have features associated with a secondary object which suggest they are either undergoing or have recently undergone a merging event with another object. Where clusters that have been classified as `relaxed' (Eq.~\ref{eq:relax}), the feature is found between $1.0-1.5~r_{500,\mathrm{true}}$ so is able to affect the mass estimate as it is within the fitting range, but does not cause the cluster to be defined as unrelaxed as this measurement was taken at $r < r_{500,\mathrm{true}}$.

Focussing on the spec masses, the clusters with $1-b < 0.5$ correspond to the clusters in Fig.~\ref{fig:GasFractions} with spec gas mass fractions that are above the universal value, $M_{\mathrm{gas,500}}/M_{500} > 0.156$. If the estimated mass of a cluster is underestimated when using the spec profiles, this will cause the spec gas mass fraction to increase such that the data point will move up and to the left with respect to the true gas mass fraction data point. The majority of the outliers are unrelaxed clusters which are typically removed from observational analyses, so it is difficult to compare these gas mass fractions to observations. There is also one cluster for CELR-E/NC in Fig.~\ref{fig:GasFractions} for which the spec gas mass fraction is noticeably underestimated for its spec mass. This cluster has $1-b > 1.5$, so its mass has been overestimated as a result of features corresponding to other objects in the gas profiles. If this additional object were masked out, it is possible that the cluster gas mass fraction might move closer to the observations. In general, the stellar mass fractions have also been increased as a result of underestimation of the more massive clusters in the samples, but the scatter is at a similar level to that between the observational samples.

\subsection{Non-thermal pressure correction}

\begin{figure*}
	\includegraphics[width=\textwidth,keepaspectratio=True]{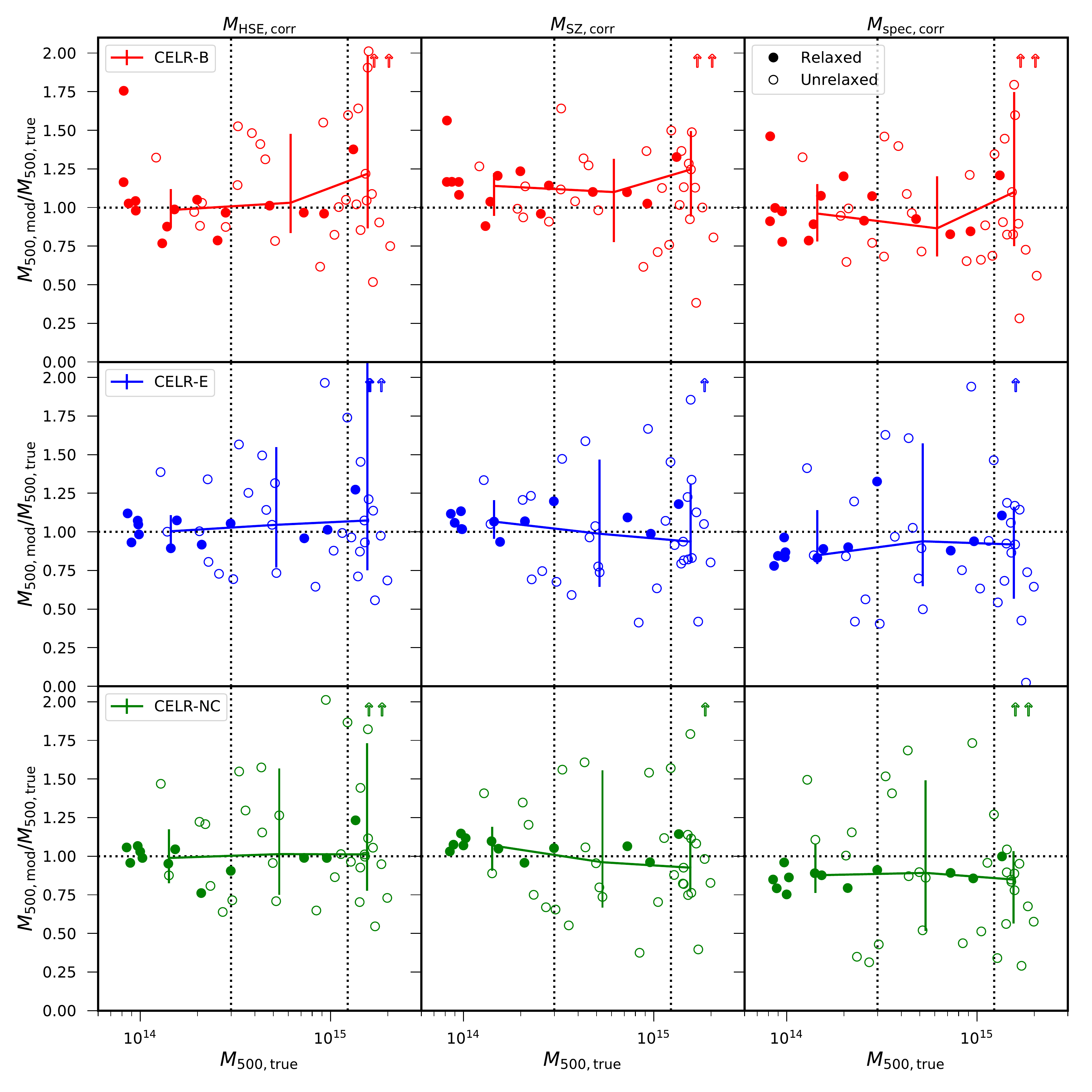}
	\caption{A measure of the offset of the estimated hydrostatic mass when including a correction for non-thermal pressure using the \citet{Nelson2014b} model compared to the true mass. The layout and style is the same as Fig.~\ref{fig:Bias}.}
	\label{fig:BiasCorr}
\end{figure*}

From Fig.~\ref{fig:PnthProfile}, non-thermal pressure becomes significant at large radii so the cluster mass estimates are expected to be underestimated compared to the true mass of the cluster without a correction for $P_{\mathrm{nth}}$. Again, assuming that the non-thermal pressure can be characterised as $P_{\mathrm{nth}}=\alpha(r)P_{\mathrm{tot}}$, then a correction to the hydrostatic mass can be applied;

\begin{equation}\label{eq:SZcorr}
M_{\mathrm{SZ,corr}} = \dfrac{1}{1-\alpha}\left[M_{\mathrm{SZ,HSE}}-\dfrac{\alpha}{1-\alpha} \dfrac{rP_{\mathrm{th}}}{G\rho_{\mathrm{gas}}}\dfrac{\mathrm{d}\ln\alpha}{\mathrm{d}\ln r}\right],
\end{equation}
\begin{equation}\label{eq:HSEcorr}
M_{\mathrm{mod,corr}} = \dfrac{1}{1-\alpha}\left[M_{\mathrm{M,HSE}}-\dfrac{\alpha}{ 1-\alpha}\dfrac{\kb Tr}{G\mu m_{\mathrm{p}}}\dfrac{\mathrm{d}\ln\alpha}{\mathrm{d}\ln r}\right].
\end{equation}
After comparing different models (see Appendix~\ref{Sec:Appendix}), the \cite{Nelson2014b} model was used to represent $\alpha(r)$ in this work. However, the results are similar regardless of the model used.

Fig.~\ref{fig:BiasCorr} shows the mass bias values for the corrected cluster masses in the same layout as Fig.~\ref{fig:Bias}. For all three runs and different mass models, the inclusion of non-thermal pressure has led to an increase in the normalisation and scatter of $1 - b$. Focussing on $M_{\mathrm{HSE,corr}}$, the median value across all mass bins and simulations is consistent within the $1\sigma$ scatter with $b=0$. As the non-thermal pressure term was introduced to correct for the assumption that clusters are in hydrostatic equilibrium, a reduction in the bias is expected.

For both $M_{\mathrm{SZ,corr}}$ and $M_{\mathrm{spec,corr}}$, including a non-thermal pressure correction helps to alleviate the mass dependence of the bias seen in Fig.~\ref{fig:Bias} for both methods. For CELR-B, the corrected SZ masses tend to be around 20 per cent above the true mass, while the CELR-E and NC simulations are consistent with no bias in all mass bias. The median mass bias is approximately constant when using spectroscopic temperature and pressure profiles with $1-b = 0.8-0.9$ for all of the runs. However, the large scatter in the bias values means any change in the distribution could again lead to a mass dependence of the bias.

Across all of the simulations and methods for mass estimation, the increase in normalisation of the mass bias when including a correction for non-thermal pressure increases with the mass of the cluster. From Eqs.~\ref{eq:SZcorr} and \ref{eq:HSEcorr}, the corrected mass estimate can be approximated by $M_{\mathrm{mod,500}}/(1-\alpha_{500})$ where $\alpha_{500}$ is the fraction of non-thermal pressure at $r_{500,\mathrm{true}}$, as the second term in both HSE equations is negligible. From Fig.~\ref{fig:PnthProfile}, $\alpha(r)$ increases with mass, hence the correction when including non-thermal pressure also increases with mass. For the lowest mass clusters this approximation is able to reproduce the median corrected mass of individual clusters to within 15 per cent and the median value to within 5 per cent. For the most massive clusters, the approximation for individual clusters drops to within 50 per cent, and the median value is reproduced to within 25 per cent. The mass estimates obtained using Eqs.~\ref{eq:SZcorr} and \ref{eq:HSEcorr} increase beyond what is expected from the simple approximation due to the value of $r_{500}$, and therefore $M_{500}$, changing due to interpolation.

\section{Summary \& future work}\label{Sec:Conclusions}

In this paper, a set of 45 clusters were simulated to investigate the effects of different subgrid models and SPH flavours on massive galaxy clusters, specifically to test the effects of these changes on the hydrostatic mass bias and a non-thermal pressure correction. The clusters were simulated using the same cosmological model as the BAHAMAS project that utilises a version of SPH which is known to lead to low entropy gas sinking to the centre of clusters due to a lack of mixing in the gas. They were also simulated using the same cosmological model used in the C-EAGLE project which has an updated hydrodynamics solver called ANARCHY that uses pressure-entropy SPH and artificial conduction to try and improve mixing at discontinuites. (The latter was turned off to create the third sample.) These different sets of runs were known as CELR-B, CELR-E and CELR-NC respectively. Analysis focused on three different mass bins: ($8.1\times 10^{13} < M_{500,\mathrm{true}}~[\mathrm{M}_{\odot}] < 3.0\times 10^{14}$); ($3.0\times 10^{14} < M_{500,\mathrm{true}}~[\mathrm{M}_{\odot}] < 1.23\times 10^{15}$); and ($M_{500,\mathrm{true}}~[\mathrm{M}_{\odot}] > 1.23\times 10^{15}$), all at redshift $z=0$. Our main results can be summarised as follows:

\begin{itemize}
	\item The CELR-B clusters have gas and stellar mass fractions which match the BAHAMAS clusters (see Figs.~5 and 7 of \citealt{McCarthy2017}) and observations (see Fig.~\ref{fig:GasFractions}). The CELR-E and CELR-NC clusters are too stellar rich across the whole mass range. This is likely due to changing from the kinetic stellar feedback model of \cite{Dalla2008} to a thermal feedback model \citep{Dalla2012} and also reducing the mass resolution of the simulation without recalibration of the feedback model. As such, thermal energy was radiated away before the gas was influenced by the increase in temperature as the sound crossing time-scale was no longer short compared to the radiative cooling timescale. However, despite not recalibrating the EAGLE model for the lower mass resolution used in this analysis, the gas and stellar mass fractions are within the scatter of the observed data, especially the relaxed subset of clusters.
	
	\item For the smallest mass clusters, CELR-E and CELR-NC are too gas rich by up to 25 per cent compared to CELR-B (see Fig.~\ref{fig:GasFractions}), likely as a result of the BHs in the EAGLE model growing slower than those using the BAHAMAS model. This resulted in central BHs which were at least a factor of 4 more massive in the CELR-B runs for all of the mass bins. Therefore, there is less energy available for AGN feedback events in the EAGLE model runs so less gas is removed from the centre of the clusters compared to CELR-B at this resolution. Although the BHs are smaller in the EAGLE model runs for all of the mass bins, the gas fractions for the runs are indistinguishable at high mass showing how changes in the subgrid physics are more evident at low mass \citep{Sembolini2016b}.
	
	\item Between the EAGLE and BAHAMAS models, the changes to the BH prescriptions leads to the BHs in CELR-E/NC being less massive than those in CELR-B. Therefore, there is less energy available for AGN feedback events in the EAGLE model runs so less gas is removed from the centre of the clusters compared to CELR-B at this resolution. This can be seen in the density profiles of the lowest mass clusters (see Fig.~\ref{fig:DensityProfile}) where the core of the CELR-B runs is less dense than the CELR-E/NC runs, and peaks at a larger radius. The BHs have displaced the gas from the centre of the cluster to the outskirts, shown by the increase in density compared to the EAGLE runs at $r > 0.8~r_{500,\mathrm{true}}$. However, there are no significant differences in the final results when separating clusters by their central BH mass.
	
	\item Within the fitting region ($0.15 - 1.5~r_{500,\mathrm{true}}$), the CELR-B and CELR-E density, temperature and pressure profiles (Fig.s~\ref{fig:DensityProfile}-\ref{fig:PressureProfile} respectively) become more comparable as the mass of the clusters increases. This shows how differences in the subgrid models are more important at low mass \citep{Sembolini2016b}. The effect of changing the SPH flavour (specifically the inclusion of artificial conduction) is most evident in the cores of the most massive clusters, shown by the median profiles in Section~\ref{Sec:GasProperties}. There was also very little difference between the gas/star/baryon fractions of CELR-E and CELR-NC whose only difference is the artificial conduction (Fig.~\ref{fig:GasFractions}).
	
	\item The fractional contribution of non-thermal pressure to total pressure for all the CELR runs continually increases beyond $0.5~r_{500,\mathrm{true}}$ in all of the mass bins (see Fig.~\ref{fig:PnthProfile}). Towards the centre of the cluster for the smallest CELR-B clusters, the fraction of non-thermal pressure also increases while the CELR-E and CELR-NC clusters continue to decrease. This is again likely due to the difference in AGN feedback prescriptions, with the EAGLE model heating only one particle per feedback event compared to 20 particles in the BAHAMAS model. As the energy is more spread out in the BAHAMAS model, this is likely to increase the outflows from the centre of the cluster and lead to a higher level of non-thermal pressure compared to the CELR-E/NC clusters.
	
	\item The result of \cite{Eckert2019}, that there is an upper limit of 13 per cent on the mean non-thermal pressure level in the \textit{Planck} clusters, is dependent on the value of $f_{\mathrm{gas,500,univ}}$ used. When using the CELR-B sample, we found that the X-COP clusters were around 20 per cent higher than our universal baryon fraction, compared to only 7 per cent when using The 300 simulations \citep{Cui2018} (Fig.~\ref{fig:DepletionFactor}). This suggests that the level of non-thermal pressure required would increase above 13 per cent.
	
	\item Looking at the hydrostatic mass biases of all the clusters (Fig.~\ref{fig:Bias}), using the spec gas temperature and density profiles increases the median bias in the most massive clusters to around 50 per cent from only 10 per cent in the smallest clusters. This is evidence for a mass dependence for the hydrostatic mass bias, confirming that seen by \cite{Henson2017} who also used mock spectroscopic data. Interestingly, the most massive clusters are biased so low compared to their true mass that they are within the regime necessary to alleviate the tension between cosmological parameters dervied from the CMB and SZ cluster counts, although drawing any firm conclusions on this would be premature as we are likely not modelling the X-ray mass estimation methods in a fully realistic way. To do this, we would need to more rigorously follow the analysis done on observed clusters such as project profiles along a line-of-sight and mask out substructures which contribute to the X-ray signals.
	
	\item Assuming that future SZ observations would have high resolution data such that it was possible to get fully resolved pressure profiles, it can be seen that the mass bias when estimating masses using pressure and density profiles reduces compared to using spec density and temperature profiles. The HSE masses are offset by around 20 per cent compared to the true cluster mass which agrees with previous simulations (e.g. \citealt{Nagai2007a, Kay2012, Rasia2012, LeBrun2014}) and highlights how the assumption of hydrostatic equilibrium is not valid in massive galaxy clusters.
	
	\item To try and reduce the bias introduced by assumed hydrostatic equilibrium (HSE) a correction was applied to the HSE equation which tried to take into account non-thermal pressure support in the clusters as a result of mergers. When the non-thermal pressure was included in the mass estimates (Fig.~\ref{fig:BiasCorr}), all the estimated masses were less biased by at least 20 per cent compared to when non-thermal pressure was not included but the scatter increased. The mass dependence of the mass bias when using mock spectroscopic data was greatly reduced when including a non-thermal pressure correction.
	
	\item The magnitude of the non-thermal pressure correction applied when estimating the cluster mass typically increases as the true mass of the cluster increases. A simple approximation for the correction gives that the corrected mass is inversely proportional to $(1-\alpha)$, where $\alpha$ is the fraction of non-thermal pressure. As previously stated, $\alpha$ typically increases with cluster mass, so the non-thermal pressure correction also increases. This simple approximation is most accurate at low masses where it is reproduces the median bias value to within 5 per cent, dropping to 25 per cent at high masses.
\end{itemize}

Our results show that, for all three samples, the hydrostatic mass bias is very similar and gives evidence for a mass dependence when using mock spectroscopic data. For the largest objects, the spec data can provide estimated cluster masses which are within the regime to reduce the tension between cosmological parameters derived from the CMB and SZ cluster counts. In the future, we will aim to analyse clusters in a way that more closely resembles the observational methods used. This will require producing more realistic mock X-ray data and a way of incorporating non-thermal pressure using line fitting. We will also aim to improve the subgrid physics models to be more realistic and include other sources of non-thermal pressure. We can then repeat our analysis using a larger cluster sample simulated at a higher mass resolution with a more robust comparison of the effects of different SPH flavours, done by incrementally changing the SPH scheme. We could also extend our analysis to compare to non-SPH methods. Currently, our results are also limited by the resolution of the simulations. It would be instructive to also repeat this analysis at higher mass resolution to better see the effects of subtle differences between the SPH methods due to mixing or the lack thereof.

\section*{Acknowledgements}
This work used the DiRAC@Durham facility managed by the Institute for Computational Cosmology on behalf of the STFC DiRAC HPC Facility (www.dirac.ac.uk). The equipment was funded by BEIS capital funding via STFC capital grants ST/K00042X/1, ST/P002293/1, ST/R002371/1 and ST/S002502/1, Durham University and STFC operations grant ST/R000832/1. DiRAC is part of the National e-Infrastructure. FAP is supported by an STFC studentship.




\bibliographystyle{mnras}
\bibliography{/Users/chessipearce/Documents/PhD/Paper/final_submission/papers.bib} 



\appendix
\section{Model fitting}\label{Sec:Appendix}

Determining the masses of clusters requires fitting analytical models to thermodynamic profiles. Throughout this analysis, profiles were obtained between $0.01-2.0~r_{500,\mathrm{true}}$ for all of the clusters by splitting the halo into 50 concentric shells and finding the median of each quantity within each radial bin. The models were then fit to these profiles within the radial range $0.15 < r_{500,\mathrm{true}} < 1.5$ using a least-squares approach with a tolerance of $1.0\times 10^{-8}$. Here, we present the true median profiles for all three samples and the corresponding fits to highlight how well the models are able to produce a smooth, parametric way of describing the data which could then be used in our analytical expressions to produce cluster mass estimates.

\subsection{Density}

The \cite{Vikhlinin2006} (heneceforth V06) model for density profiles is an extension of the $\beta$-model proposed by \cite{Cavaliere1978}. In this analysis, the additional term in the full V06 model used to increase the fitting ability of cluster cores, $r \lesssim 0.1~r_{500}$, was not included as the fitting range adopted here was $0.15 - 1.5~r_{500,\mathrm{true}}$. Therefore, the model used was
\begin{equation}
n_{\mathrm{p}}n_{\mathrm{e}} = n_0^2 \frac{(r/r_{\mathrm{c}})^{-\alpha}}{(1+r^2/r_{\mathrm{c}}^2)^{3\beta-\alpha/2}}\frac{1}{(1+r^{\gamma}/r_{\mathrm{s}}^{\gamma})^{\epsilon/\gamma}},
\end{equation}
where $n_{\mathrm{p}}$ is the proton density, $n_{\mathrm{e}}$ is the electron density, $n_0$ is a normalization constant, $r_{\mathrm{c}}$, $r_{\mathrm{s}}$ are scale radii, and $\alpha$, $\beta$, $\gamma$, $\epsilon$ control the slope and width of the transition regions. To stop unphysically sharp density peaks V06 imposed a constraint of $\epsilon < 5$, and suggest that fixing $\gamma = 3$ leads to a good fit. This resulted in a model with 6 free paramters.

\begin{figure}
	\includegraphics[width=\columnwidth,keepaspectratio=True]{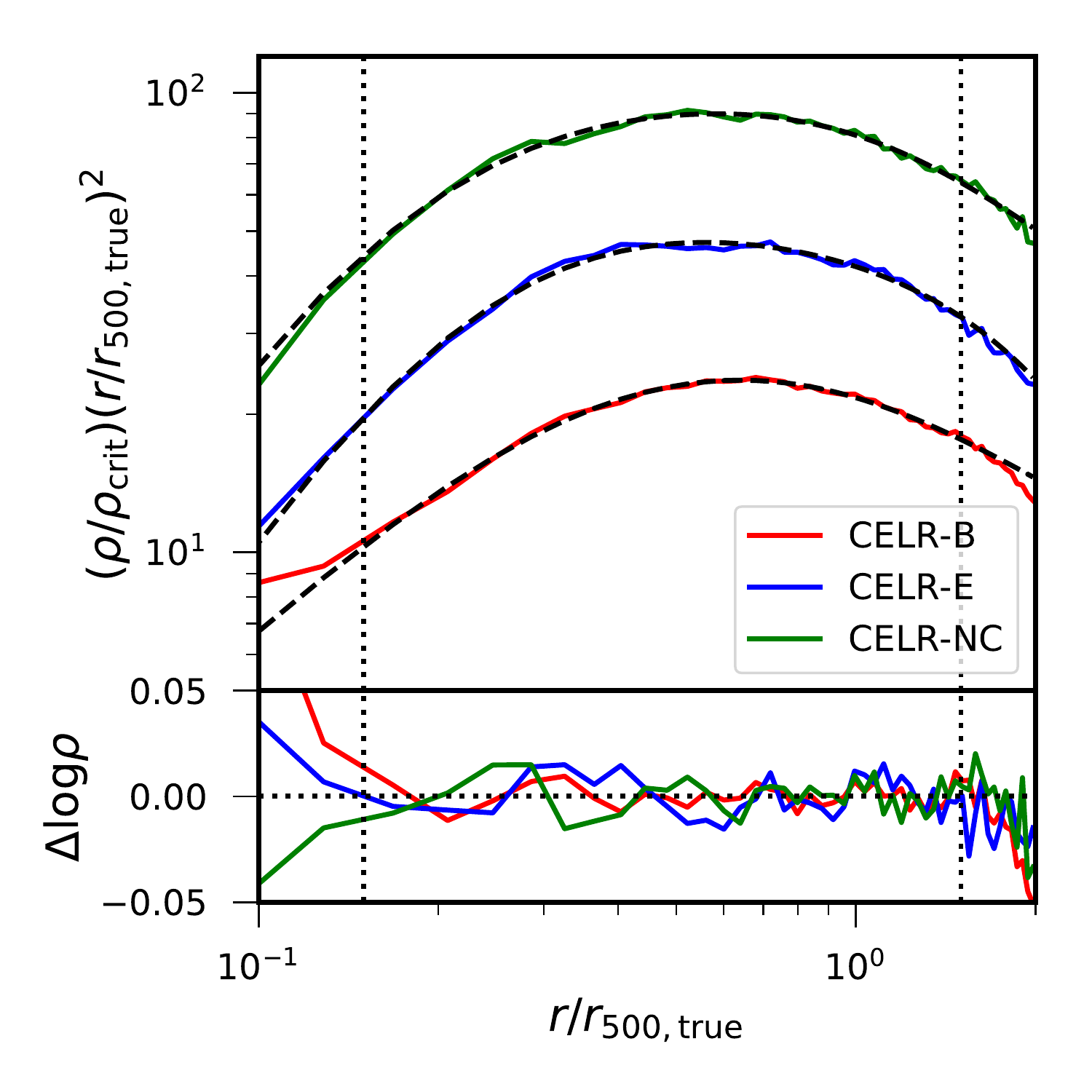}
	\caption{The true density profiles for clusters with mass $3.0~\times~10^{14}~< M_{\mathrm{true}}~[$\msol$] < 1.23\times 10^{15}$ for CELR-B (red), CELR-E (blue) and CELR-NC (green). The CELR-E and CELR-NC profiles are offset by a factors of 2 and 4 respectively. The vertical dotted lines show the fitting region for the \citet{Vikhlinin2009} six parameter density model shown by the black dashed lines for all of the profiles.}
	\label{fig:DensityFit}
\end{figure}

Fig.~\ref{fig:DensityFit} shows the median density profiles for the middle mass bin, $3.0\times 10^{14} < M_{\mathrm{true}}~[$\msol$] < 1.23\times 10^{15}$, for all of the cluster samples. CELR-E (blue line) and CELR-NC (green line) have been offset by aribtrary factors of 2 and 4 respectively from their actual densities, and the dashed black lines show the Vikhlinin model using the best fit parameters to each of the profiles. From the residuals (bottom panel), the V06 model is able to recreate the median density profiles of all of the CELR samples to within 5 per cent across the full fitting range (shown by the verticle dotted lines in both panels).

\subsection{Temperature}

V06 also proposed a model to fit temperature profiles;
\begin{equation}
T(r) = T_0\frac{(r/r_{\mathrm{cool}})^{a_{\mathrm{cool}}}+(T_{\mathrm{min}}/T_0)}{(r/r_{\mathrm{cool}})^{a_{\mathrm{cool}}}+1}\frac{(r/r_t)^{-a}}{(1+r^b/r_t^b)^{c/b}}
\end{equation}
where $T_0$, $T_{\mathrm{min}}$ are normalization constants, $r_{\mathrm{cool}}$, $r_{\mathrm{t}}$ are scale radii, and $a_{\mathrm{cool}}$, $a$, $b$, $c$ control the slope and width of the transition regions. None of the parameters were kept constant leading to an eight parameter fit.

\begin{figure}
	\includegraphics[width=\columnwidth,keepaspectratio=True]{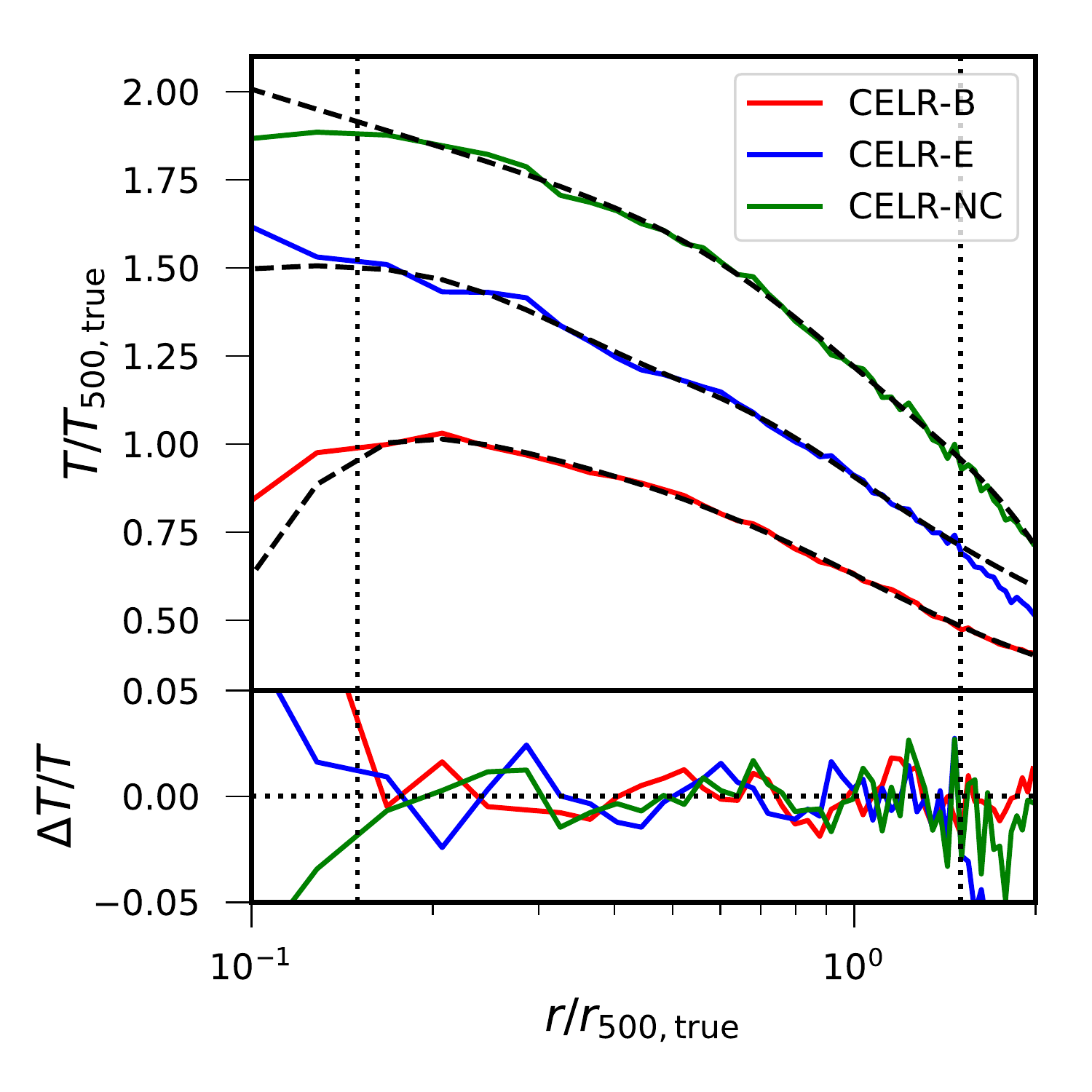}
	\caption{The \citet{Vikhlinin2006} eight parameter temperature model fits to the true median profiles for all of the samples in the same mass range and using the same colour scheme as Fig.~\ref{fig:DensityFit}. Again, CELR-E and CELR-NC are offset by arbitrary factors of 1.5 and 2 respectively.}
	\label{fig:TemperatureFit}
\end{figure}

The true temperature profiles for the middle mass bin are shown in Fig.~\ref{fig:TemperatureFit}. CELR-E and CELR-NC have been offset by an arbitrary factor of 1.5 and 2 respectively. The linestyles and colours are the same as Fig.~\ref{fig:DensityFit}. From the residuals, the V06 temperature model is able to reproduce the true profile to within 5 per cent over the whole fitting range for all of the samples. In general, there is less scatter in the fit to the temperature profile than there is for the density profiles.

\subsection{Thermal pressure}

To fit to the pressure profiles, the generalised NFW (GNFW) model introduced by \cite{Nagai2007b} was used;
\begin{equation}
\dfrac{P(r)}{P_{500}} = \dfrac{P_0}{(r/r_{s})^{\gamma}[1+(r/r_{s})^{\alpha}]^{(\beta-\gamma)/\alpha}}
\end{equation}
where $P_0$ is a normalization constant, $r_{s}$ is the scale radius, and ($\alpha$, $\beta$, $\gamma$) are parameters that control the slope at $r \sim r_{s}$, $r \gg r_{s}$ and $r \ll r_{s}$ respectively. The value $\gamma = 0.31$ was adopted from the universal pressure profile of \cite{Arnaud2010} due to the degeneracy between $\beta$ and $\gamma$, so the final model was a four parameter fit.

\begin{figure}
	\includegraphics[width=\columnwidth,keepaspectratio=True]{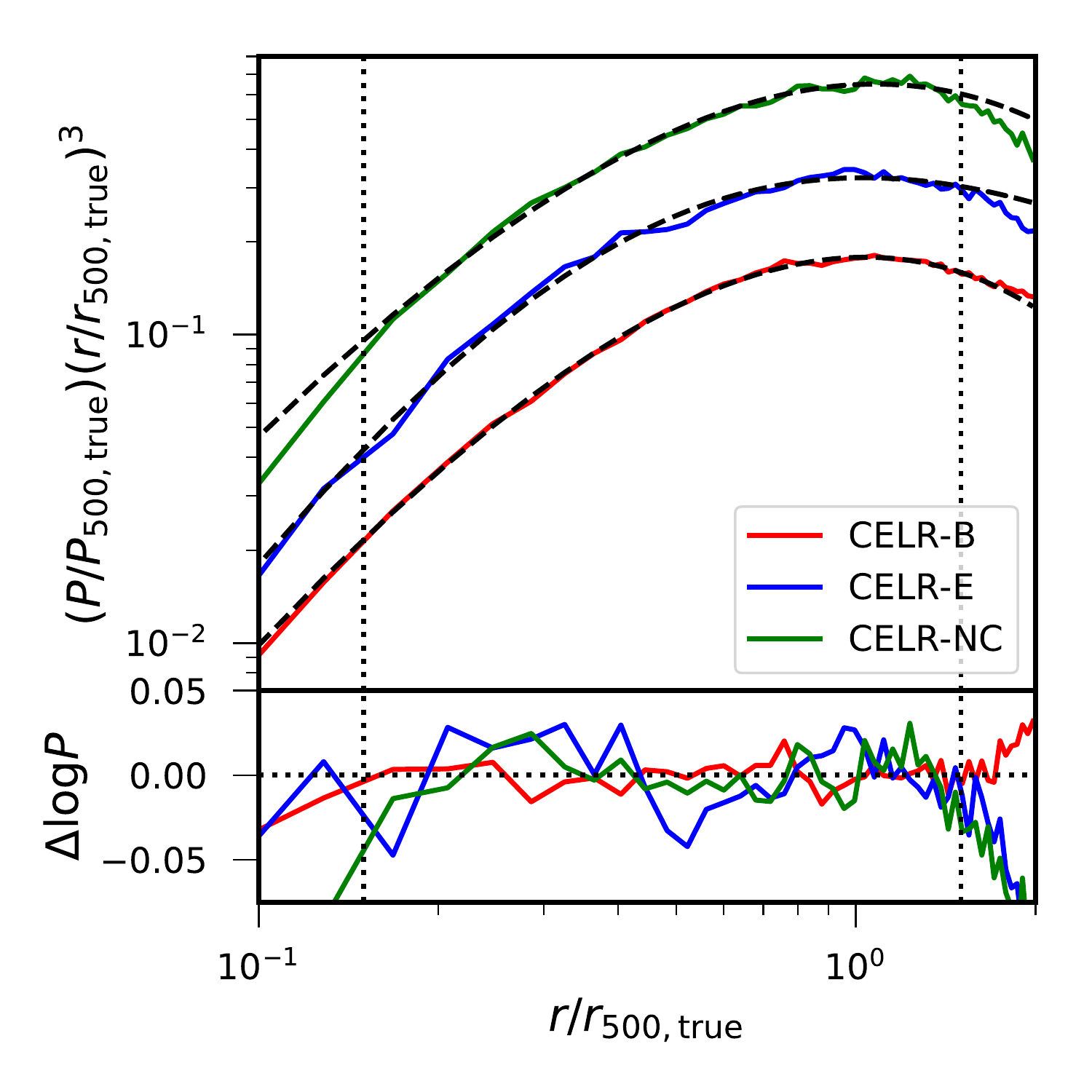}
	\caption{Fits to the true median pressure profiles using the four parameter generalised NFW model (e.g. \citealt{Navarro1996, Nagai2007b, Arnaud2010}) following the same layout and colour scheme as Fig.~\ref{fig:DensityFit}. The CELR-E and CELR-NC are offset by a factor of 2 and 4 respectively. The GNFW model is able to recreate the pressure profiles to within 20 per cent depending on the initial sample.}
	\label{fig:PressureFit}
\end{figure}

Fig.~\ref{fig:PressureFit} shows the fit of the GNFW model to the true median profiles of the middle mass bin follwing the same linestyle and colour scheme as Fig.~\ref{fig:DensityFit}. The CELR-E and CELR-NC profiles have been offset by a factor of 2 and 4 respectively. From the resdiuals, the GNFW model is able to produce smooth pressure profiles within 20 per cent of the true profile for the whole of the fitting range. 

\subsection{Non-thermal pressure}

Three different fitting models for the non-thermal pressure (written as $P_{\mathrm{nth}} = \alpha P_{\mathrm{th}}$) were tested. The first is from \cite{Nelson2014b} which suggested the form
\begin{equation}
\alpha(r) = 1 -  A\left(1+\exp\left[-\left(\dfrac{r/r_{500}}{B}\right)^{C}\right]\right),
\end{equation}
where $A$, $B$ and $C$ are the three free parameters. This is slightly different to the model used by \cite{Nelson2014b} who use $r_{200m}$, the radius which encloses 200 times the mean density of the Universe, instead of $r_{500}$. When fitting, the initial guess for the parameters were taken as the best fitting values from \cite{Nelson2014b}, $A = 0.45$, $B = 0.84$ and $C = 1.63$.

The second model is from \cite{Fusco2014} which assumes that the non-thermal pressure decays on the scale of $B$ inward,
\begin{equation}
\alpha(r) = A\exp\left[-\left(\dfrac{1-r/(2r_{500})}{B}\right)^2\right],
\end{equation}
where $A$ and $B$ are the two free parameters.

Finally, \cite{Shaw2010}  suggested a model which assumes a simple power law with cluster centric radius,
\begin{equation}
\alpha(r) = A\left(\dfrac{r}{r_{500}}\right)^C,
\end{equation}
where $A$ and $C$ are the two free parameters. When fitting, the inital guess for the parameters was taken as the best fitting values from \cite{Shaw2010}; $A = 0.18$ and $C = 0.8$.

\begin{figure}
	\includegraphics[width=\columnwidth,keepaspectratio=True]{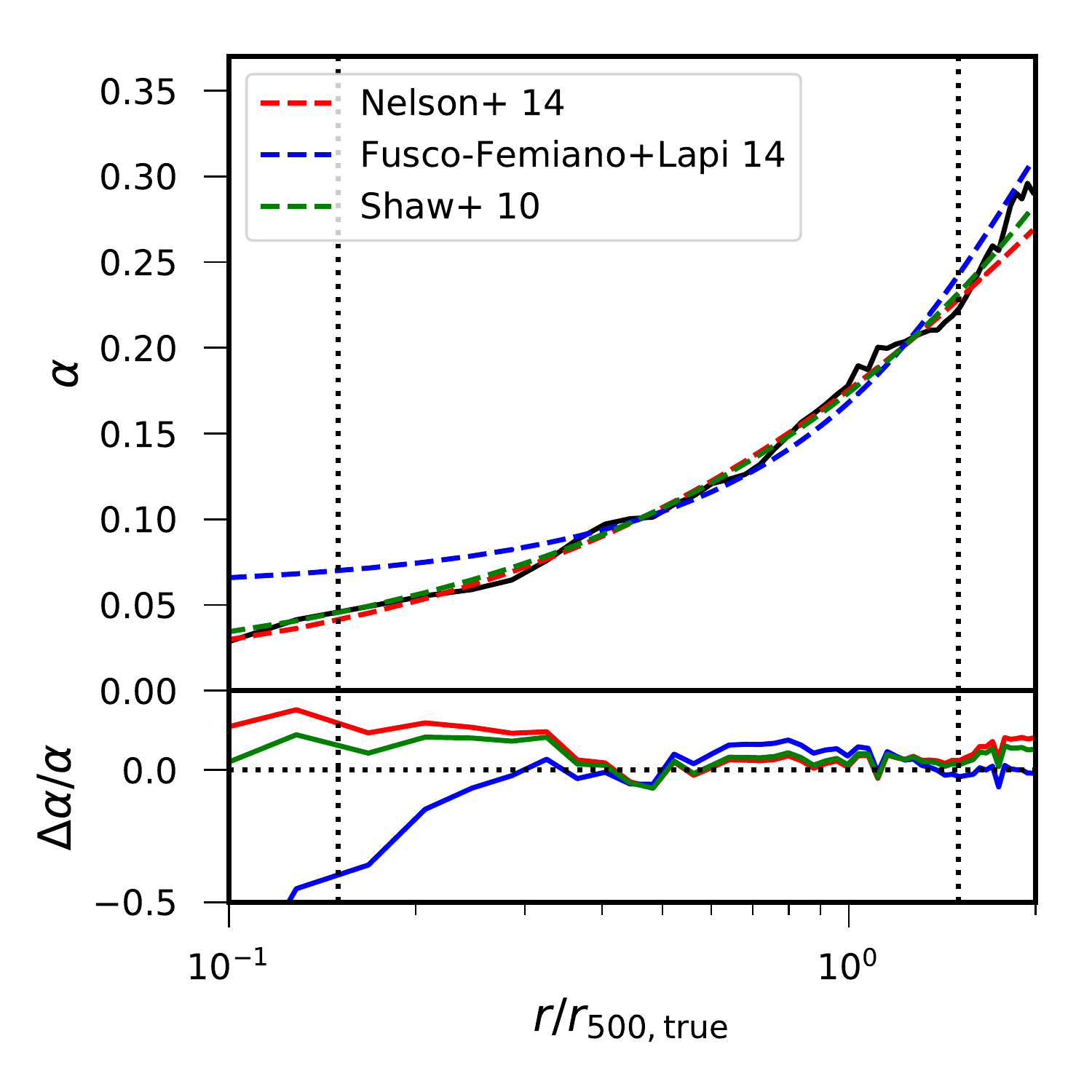}
	\caption{Fits to the median non-thermal pressure profiles of the middle mass clusters of the CELR-E samples using the \citet{Nelson2014b} model (red line), \citet{Fusco2014} model (blue line) and the \citet{Shaw2010} model (green line).}
	\label{fig:PnthFit}
\end{figure}

Fig.~\ref{fig:PnthFit} shows the result of all three of the non-thermal pressure models fitting to the median $\alpha(r)$ profile for the CELR-E clusters in the middle mass bin. The \cite{Nelson2014b} model is shown in red, \cite{Fusco2014} model is blue and the \cite{Shaw2010} model is green. From the residuals, the \cite{Nelson2014b} and \cite{Shaw2010} models are both able to reproduce the CELR-E profile to within 10 per cent while the \cite{Fusco2014} model is less able to follow the curve. The \cite{Nelson2014b} model was chosen to represent the non-thermal pressure profiles when adding a correction to the hydrostatic mass due to it being better able to reproduce individual cluster $\alpha(r)$ profiles.


\bsp	
\label{lastpage}
\end{document}